\documentclass[%reprint,
nofootinbib,
 amsmath,amssymb,
]{revtex4-1}
\usepackage{epsf}
\usepackage{multirow}
\usepackage{bm}% bold math
\usepackage{color}

\usepackage[dvipsnames]{xcolor}
\usepackage{graphicx}
\usepackage{epstopdf}
\usepackage{float}
\def\MSbar{{\rm  \overline{\footnotesize MS\kern-0.05em}\kern0.05em}}

\newcommand{\Dlr}{\buildrel \leftrightarrow \over D\raise-1pt\hbox{}\!\!}
 \newcommand{\Dl}{\buildrel \leftarrow \over D\raise-1pt\hbox{}}
\newcommand{\Dr}{\buildrel \rightarrow \over D\raise-1pt\hbox{}}

\newcommand{\be}{\begin{eqnarray}}
\newcommand{\ee}{\end{eqnarray}}

\newcommand\defeq{\stackrel{\textup{\tiny def}}{=}}

\begin{document}

\title{Markov Chain Monte Carlo technics applied to Parton Distribution Functions determination: proof of concept} 

\author{Y\'emalin Gabin Gbedo, Mariane
Mangin-Brinet
}
\affiliation{%
Laboratoire de Physique Subatomique et de Cosmologie - Universit\'e Grenoble-Alpes, CNRS/IN2P3, 
53, avenue des Martyrs, 38026 Grenoble, France\\
}%

%\date{\today}

%\numberwithin{equation}{section} 

\preprint{LPSC????}

%%%%%%%%%%%%%%%%%%%%%%%%%%%%%%%%
% abstract
%%%%%%%%%%%%%%%%%%%%%%%%%%%%%%%%
\begin{abstract}

We present a new procedure to determine Parton Distribution Functions (PDFs), based on Markov Chain Monte Carlo (MCMC)
methods. The aim of this paper is to show that we can replace the standard $\chi^2$ minimization by procedures grounded on Statistical Methods, and on Bayesian inference in particular, thus offering additional insight into the rich field of
PDFs determination.  
After a basic introduction to these technics, we introduce the algorithm we have chosen to implement -- namely Hybrid (or Hamiltonian) Monte Carlo. This algorithm, initially developed for Lattice QCD, turns out to be very interesting when applied to PDFs determination by global analyses; we show that it allows to circumvent the difficulties due to the high dimensionality of the problem, in particular concerning
the acceptance. A first feasibility study is performed and presented, which indicates that Markov Chain Monte Carlo can successfully be applied to the extraction of PDFs and of their uncertainties. 
\end{abstract}
%%%%%%%%%%%%%%%%%%%%%%%%%%%%%%%%
% insert suggested PACS numbers in braces on next line
%\pacs{}
%Key words:  Statistical analysis; Markov Chain Monte Carlo; PDFs; structure function; QCD; Hybrid Monte Carlo

\maketitle

%\begin{flushleft}
%DAMTP-2010-??\\
%LPT-Orsay 10-??\\
%UHU-FT/10-?? \\
%LPSC10122
%\end{flushleft}

%\tableofcontents

%%%%%%%%%%%%%%%%%%%%%%%%%%%%%%%%%%%%%%%%%%%%%%%%%%%%%%%%
\tableofcontents
\newpage
\section{Introduction and motivation}

Quantum Chromodynamics (QCD) is the theory of strong interaction, whose ambition is to explain nuclei cohesion as well as neutron and proton structure, i.e. most of the visible matter in the Universe.  Its application domain is even wider, since QCD controls the structure and interactions of all hadrons: proton, neutron, hyperons, pions, kaons,...It is one of the most elegant theory of Science History (with General Relativity); it has only very few parameters and allows to give a physical interpretation to a very broad range of phenomena using a well defined and very compact formalism.

Among the fundamental ingredients of QCD, Parton Distribution Functions (PDFs) are key elements 
and play an essential role to connect the QCD dynamics of quarks and gluons to the measured hard scattering cross sections of colliding hadron(s).
They carry an invaluable source of information on hadrons partonic structure and enormous theoretical 
and experimental efforts since years have been devoted to the extraction of these distribution functions.

PDFs are all the more important nowadays that, with the start of data taking at the LHC, they are essential for
the computation of a large class of observables. Built for the discovery of the Higgs
boson and the study of physics beyond Standard Model, the LHC is indeed essentially
a îQCD factoryî, producing events in an unexplored energy range. The LHC potential
of discovery crucially depends on the quality of predictions for QCD signals and
backgrounds and thus on the PDFs quality.

PDFs are intrinsically non-perturbative objects and thus cannot be determined using only perturbative QCD tools. One of the most efficient method to 
perform non perturbative QCD calculations is Lattice QCD. However, although 
nucleon structure has been the subject of intensive activity in Lattice QCD since years and even if promising technics to compute PDFs 
directly on the lattice have recently being proposed \cite{PDF_LQCD}, {\it{ab initio}} calculations of PDFs are very
challenging and still not a competitive alternative to global analyses. 
These latter thus remain the chosen method to obtain PDFs, which are parametrized by functional forms whose parameters are constrained 
by fits to the data. 

Parton Distribution Function determination thus consists in extensive exploitation of datasets collected at colliders to constrain the parameters of the PDF functional forms given at a fixed scale in energy. Such analyses are usually based on a least square fit method, i.e. on the minimization of a $\chi^2$, 
which compares the input data and theory predictions. 
%constructed using the experimental cross-sections and corresponding theoretical calculations. 
PDFs determined this way did not for a long time include any estimate of uncertainties, other than the mere comparison of results provided by different global analyses collaborations. However, with the advent, at the dawn of the 21st century, of
the new generation of colliders and the active search for New Physics, the need
to assess the uncertainties of the PDFs became clear \cite{Lai_PRD97,Martin_DTP96,Giele_98,Huston_96}.  Many studies have since been devoted to the estimates of uncertainties on physical predictions due to the uncertainties of PDFs (see for instance \cite{J_Pumplin_PRD2001, D_Stump_PRD2001,Giele_2001} and references therein) and considerable progress has been made. Nevertheless, this task is far from being trivial and many issues remain open \cite{J_Pumplin_PRD2001}.  
%Sources of uncertainties, correlated or uncorrelated, are both experimental and
%theoretical and their analysis is far from being a trivial task.

Our current understanding of PDFs uncertainties is mainly based -- however with the notable exception of neural network technics --  on the Hessian method or the Lagrange multiplier one \cite{J_Pumplin_PRD2001, D_Stump_PRD2001}. The determination of the uncertainties then relies on an assumption on the permissible
range of "acceptable" $\Delta\chi^2$ for the fit and the choice of a tolerance parameter. 
In order to improve on this point and get a deeper insight, we propose to use Markov Chain Monte Carlo (MCMC) techniques to define the uncertainties in a way 
based as much as possible on robust statistical methods. 
%a more rigorous manner. 
Markov Chain Monte Carlo algorithms have been an extremely popular tool in statistics. 
While these techniques are already widely used in many areas of physics (see for instance \cite{Gilks,Sokal_1989}) they have not yet been employed 
as a standalone method to compute PDFs parameters and their errors -- i.e. without resorting to a $\chi^2$ minimization procedure\footnote{Uncertainties estimations using pseudodata replicas are also based on Monte Carlo methods \cite{Giele_2001}, but still rely on fits.}. 
%replicas }. 
%in the context of determination of parton distribution functions. 

% Mention Monte Carlo replicas of data

%The Monte Carlo (MC) technique
%[104, 105] can also be used to determine PDF uncertainties.
%The uncertainties are estimated using pseudodata
%replicas (typically >100) randomly generated from
%the measurement central values and their systematic and
%statistical uncertainties taking into account all point-topoint
%correlations. The QCD fit is performed for each replica and the PDF central values and their experimental
%uncertainties are estimated from the distribution of
%the PDF parameters obtained in these fits, by taking the
%mean values and standard deviations over the replicas.
%arXiv:1410.4412v3

%DIFFICULTE DE LA MINIMISATION STANDARD - nombre de param\`etres, et surtout incertitudes. \par
%POSSIBILITE D'IDENTIFIER LES DIRECTIONS PLATES ET LES MINIMA LOCAUX. \par

%There have been many developments in recent years beyond the conventional analyses that underlie the widely used PDFs. These developments have been driven by the need to quantify the uncertainties of the PDFs and their physical predictions. The problem of estimating uncertainties of the PDFs is usually addressed by examining alternative fits obtained by subjective tuning of selected degrees of freedom. The goal of this work is to find a way to assess the uncertainties objectively, applying Markov Chain Monte Carlo techniques to global analyses.

The Markov Chain Monte Carlo method allows to estimate {\it{a posteriori}} probability densities for multi-dimensional models and provides reliable estimates of errors. 
%MCMC algorithms enable us to draw samples from a probability distribution known up to a multiplicative constant. 
MCMC consist in sequentially simulating a single Markov chain whose limit distribution is the chosen one.

%If successful, this approach has the potential to advance the entire field of global PDFs analyses and hence of pQCD. 
The main challenge of the present paper is to demonstrate that Markov Chain Monte Carlo technics can be applied 
to PDFs extraction. The higher the dimension of the chain (i.e. in our case, the more PDFs free parameters to be determined), the
more computing time is needed to generate the chain. The large number of parameters to be computed in a full PDFs determination 
has led us  to make use of a Monte Carlo algorithm based on molecular dynamics, initially developed for lattice field theory. 
We apply this algorithm to a realistic (though not full) extraction of PDFs, based on 10-parameter functional forms and four
data sets, to demonstrate that Markov Chain Monte Carlo can successfully be applied to PDFs computation.

% parler de la possibilit? de fitter alpha_S en intro, ou seulement en conclusion? (ca depend si on a des r?sultats pour alpha_S)

This paper is organized as follows: in section \ref{sec:Bayes}, we formulate the PDFs determination problem in terms of Bayesian
inference. In the next section, the basic principles of Markov Chain Monte Carlo methods are recalled and
illustrated utilizing the widely used Metropolis algorithm. 
%with one dimensional examples obtained by setting all PDFs parameters but one to their values obtained after standard $\chi^2$ minimization. 
The fourth section briefly presents the Hybrid Monte Carlo algorithm and shows how it allows to deal with the
large number of PDFs parameters to be determined. Section \ref{sec:Analysis} details the MCMC analysis procedure and section \ref{sec:Results} displays first results with a realistic run using 10 parameters and 4 sets of data. Conclusions and outlook are discussed in the last section.

%%%%%%%%%%%%%%%%%%%% Metropolis %%%%%%%%%%%%%%%%%%
%\section{Formulation of PDFs determination in terms of Markov Chain Monte Carlo} 
\section{Formulation of the problem in terms of Bayesian inference}\label{sec:Bayes} 

Parton Distribution Functions are usually (with the exception of Neural Network procedure \cite{DelDebbio:2007ee}) parametrized at a given energy scale by functional forms, which are then evolved at any other scale thanks the DGLAP equations (this also excepts dipole models \cite{Nikolaev_1991} and transverse-momentum dependent \cite{Collins_2011} and unintegrated PDFs \cite{Aybat_2011} for instance, that we will not consider here). These PDFs are convoluted with partonic cross-sections to obtain hadronic cross-sections for various 
processes and a $\chi^2$ function, constructed from these theoretical cross-sections and corresponding experimental data, is then minimized to constrain the PDFs parameters. Rather than using a minimization procedure and a Hessian method to estimate PDFs uncertainties, we propose a Bayesian parameter inference approach. These technics have already been successfully applied in many areas \cite{Gilks} and we only sketch the main principles in what follows. The interested reader can referred for instance to \cite{MCMC_Neal} for a more extensive review of the subject. 

For compactness, we note $\hat{q}$ the vector
of PDFs parameters to be determined: $\hat{q}=(q^{(1)},q^{(2)},\ldots,q^{(m)})^T$ where $m$ is typically, in the case of a full analysis, of the order of 25-30, and $D$ the data. From a Bayesian perspective, both model parameters and observables are considered random quantities, and Bayesian inference aims formally to determine a joint probability distribution $P(D,\hat{q})$ over all random quantities. This joint distribution can be written
as $P(D,\hat{q})=P(D|\hat{q})P(\hat{q})$, where $P(\hat{q})$ is a {\it{prior}} distribution -- quantifying the degree of belief one has {\it{a priori}}
before observing the data -- and $P(D|\hat{q})$ is the likelihood of the data: $\mathcal{L}(\hat{q})\defeq P(D|\hat{q})$. 
Bayes theorem is used to express the distribution of $\hat{q}$ conditional on $D$, $P(\hat{q}|D)$, in terms of the likelihood $P(D|\hat{q})$:
\begin{eqnarray}\label{eq:PthetaD}
P(\hat{q}|D)=\frac{P(D|\hat{q})P(\hat{q})}{\int\,d\hat{q}P(D|\hat{q})P(\hat{q})}
\end{eqnarray}  
The denominator in (\ref{eq:PthetaD}) does not depend on the parameters and can be considered only as a normalization. 
This so-called {\it{posterior}} probability density $P(\hat{q}|D)$ quantifies the probability to have the model parameters $\hat{q}$ given the observed data $D$ and
is the object we deal with in all Bayesian inference. To determine this conditional probability, we thus need to set a prior distribution for the parameters,
and to compute the likelihood of the data. The probability density $P(\hat{q}|D)$ is then sampled using a Monte Carlo algorithm.  

%The probability density $P(\hat{q}|D)$ is then generated by a Monte Carlo algorithm.  

Assuming that the fluctuations of the $n$ experimental data points under consideration around their corresponding theoretical values are uncorrelated and distributed according to a Gaussian law (assumption whose validity can be assessed {\it{a posteriori}} - see section \ref{sec:Results}), the least square method 
and the maximum likelihood are equivalent and the logarithmic likelihood function can be written as 
\begin{eqnarray}\label{eq:Like}
\log \mathcal{L}(\hat{q})=-\frac{1}{2}\displaystyle\sum_{i=1}^{n}\frac{(D_i-T_i)^2}{\sigma_i^2}=-\frac{1}{2}\chi^2
\end{eqnarray}  
where $D_i$ and $T_i$ denote respectively the $i^{th}$ experimental point and the corresponding theoretical calculation, and $\sigma^2_i$ is the uncertainty associated with the measured data $i$.
% Changer la notation pour eviter les deux utilisations de sigma?
The inclusive cross-section $T_i$ in hadron collision can be written as a convolution of PDFs with a partonic cross-section, computed at a given order in perturbation theory. The likelihood function (\ref{eq:Like}) thus contains the PDFs. 
Correlated experimental uncertainties can also be taken into account by introducing for instance a covariant matrix and properly modifying the $\chi^2$ \cite{HeraFitter}.

For this work, we have used the $\chi^2$ function given by the default settings provided by the HeraFitter package \cite{HeraFitter} with 10 parameters (default HeraFitter steering file and minuit.in.txt.10pHERAPDF input file) and four data sets with an initial PDFs parametrization
set at $Q^2_0=1.9$ GeV$^2$ and a lower cut on the data at $\mathrm{Q}_{min}^2 = 10$ GeV$^2$ . 
These settings provide already a computation of PDFs realistic enough for this feasibility study. 

%CPU time CHALLENGE?
%\bigskip  

We thus apply Bayesian inference to the likelihood function defined in (\ref{eq:Like}), that is, we compute the probability density function of the model
parameters, based on selected experimental data. To this purpose, we use Monte Carlo Markov Chain procedure, whose principles are briefly sketched in the next section. One of the crucial interest of these methods is the fact that the mean value and uncertainty in these parameters are by-products of the probability density functions computed.

%including the information on their uncorrelated and correlated uncertainties
% Correlated experimental uncertainties can be accounted for using a nuisance
% parameter method or a covariance matrix method. Different statistical assumptions for
%the distributions of the systematic uncertainties, e.g. Gaussian
%or LogNormal, can also be studied

%%%%%%%%%%%%%%%%%%%% Metropolis %%%%%%%%%%%%%%%%%%
\section{Principle of Markov Chain Monte Carlo} 

%Markov Chain Monte Carlo methods are based on Bayesian statistics, whose principle -- as stated in the previous section -- is the determination of individual posterior probability density function.

\subsection{Basics of the method}
%(Gilks, Richardson and Spiegelhalter 1996, Robert and Casella 1999).

The Markov Chain Monte Carlo method allows to estimate {\it{a posteriori}} probability densities for multi-dimensional models -- which, as explained briefly in the previous section, is exactly what we want -- and provides reliable estimates of errors. MCMC algorithms enable us to draw samples from a probability distribution known up to a multiplicative constant. They consist in sequentially simulating a single Markov chain whose limiting distribution is the chosen one (in our case, the maximum likelihood times a prior density). 
More precisely, a Markov Chain is a stochastic process characterized by the fact that the conditional
distribution of the random variable at iteration $t$, denoted $\hat{q}_{t}$, given the ensemble of random variables at all previous steps
$\hat{q}_0,\ldots ,\hat{q}_{t-1}$, depends only on $\hat{q}_{t-1}$, and not on the previous history. 
Such a chain can be used to sample a probability density. To converge to a given
stationary distribution, the chain needs to satisfy important properties: it has to be irreducible, aperiodic and positive recurrent. We will
not expand further on Markov chain theory and we refer the reader interested by formal details to \cite{Gilks} and references therein.

Two ingredients are necessary to define a Markov Chain: (i) the initial values (that is the marginal distribution) of parameters and (ii) the transition
kernel between two sets of parameters: $T(\hat{q}\longrightarrow\hat{q}')$, for going from a set $\hat{q}$ to another set
$\hat{q}'$. There are several issues arising when implementing MCMC: the influence of the starting point of the chain (leading to the "burn-in" time),
the choice of the transition kernel, the rate of convergence, the acceptance of the algorithm,\ldots. These questions will be illustrated in
detail in the following sections.

\subsection{Metropolis algorithm}

The so called "Metropolis-Hastings algorithm", proposed in 1953 by Metropolis et al.\cite{Metropolis} and generalized by Hastings in 1970 \cite{Hastings}, is one of the simplest Monte Carlo algorithm. It is the standard computational workhorse of MCMC methods both for its simplicity and its versatility, and is in principle applicable to any system. 
It is extremely straightforward to implement and to sample a target density $P(\hat{q}|D)$ (see section \ref{sec:Bayes}) it
proceeds as follows: at each Monte Carlo time $t-1$, the next state $\hat{q}_{t}$ is chosen by sampling a candidate point $\hat{q}'$ from a proposal distribution $\pi(.|\hat{q}_{t-1})$. The candidate point is then accepted with the probability 
$$
\alpha(\hat{q}_{t-1},\hat{q}')=\min\Big(1,\frac{P(\hat{q}'|D)\pi(\hat{q}_{t-1}|\hat{q}')}  {P(\hat{q}_{t-1}|D)
\pi(\hat{q}'|\hat{q}_{t-1})}\Big)
$$
and the Metropolis-Hastings transition kernel is thus 
$$
T(\hat{q}_{t-1}\longrightarrow\hat{q}')=\pi(\hat{q}'|\hat{q}_{t-1})\alpha(\hat{q}_{t-1},\hat{q}').
$$
If the new set of parameters $\hat{q}'$ is accepted, the next state of the chain becomes $\hat{q}_{t}=\hat{q}'$. If it is rejected, the chain does not move and the point at $t$ is identical to the point at $t-1$: $\hat{q}_{t}=\hat{q}_{t-1}$.

\bigskip

A special case of the Metropolis-Hastings algorithm is the random walk Metropolis, for which the proposal distribution is chosen to
be such that $\pi(\hat{q}'|\hat{q}_{t-1})=\pi(|\hat{q}_{t-1}-\hat{q}'|)$. The acceptance probability then reduces to $\alpha(\hat{q}_t,\hat{q}')=\min\Big(1,\frac{P(\hat{q}'|D)}  {P(\hat{q}_t|D)}\Big)$. Frequently, the proposal for the random walk jump has a form which depends
on a scale parameter, giving the typical "size" of the leap from one site to the other. For instance the proposal distribution for $\hat{q}'$ could be a normal distribution centered in $\hat{q}_{t-1}$ with a standard deviation $\sigma$. A meticulous attention has to be taken when choosing this
scale parameter. If it is too large, a very high percentage of the iterations will be rejected, leading to an inefficient algorithm. If it is
too small, the random walk will explore the parameter space very slowly, leading again to inefficiency. This problem is all the more difficult
to handle that the number of parameters (i.e. the dimension of the vector $\hat{q}$) to be sampled increases. 

\bigskip
Ideally, to optimize the efficiency of the MCMC, the proposal distribution should be as close as possible to the target distribution. In practice, 
the performance of the algorithm is obviously highly dependent on the choice of proposal distribution $\pi(.|\hat{q}_{t-1})$ and 
several options are usually considered in the literature to explore the parameter space: one-dimensional Gaussian distributions, multivariate Gaussian distributions, a distribution obtained by binary space partitioning \cite{deBerg_2000} \ldots. However, even if adjustments of the proposal distributions improve the Metropolis efficiency, they are not effective enough to efficiently deal with several dozen parameters within a reasonable CPU and user time. To circumvent these problems -- 
%Actually, this kind of Metropolis algorithm, where proposal distributions need to be tuned, turns out to be inapplicable when dealing with more than 10-15 parameters. 
since in the case of PDFs extraction, the number of free parameters to determine (that
is the number of parameters in the PDFs functional form) is of the order of $\sim25-30$ -- we have implemented a much more efficient algorithm, based on Molecular Dynamics, which has initially been developed for Lattice QCD and is widely used in this field. 

%a much more efficient algorithm (see section \ref{sec:HMC}) and to keep Metropolis algorithm only
%to illustrate main features of MCMC. 

%As a rule of thumb, the step from one point to the next proposed one is adjusted empirically so
%that the acceptance ratio for one parameter is in the range 30-50\%, with 30-40\% tending to work better than 40-50\% (REFERENCE). 

%However,  as mentioned previously, the acceptance drops drastically when increasing the number of parameters.  

%%%%%%%% HMC %%%%%%%%%%%%%%%%%
%%%%%%%% HMC %%%%%%%%%%%%%%%%%
\section{Hybrid (or Hamiltonian) Monte Carlo}\label{sec:HMC}

As mentioned before, the main problem of Metropolis type algorithms, relies on the choice of the candidate point at each move of the chain. Choosing
a trial point far from the initial one will lead to a large change in the distribution to sample, and thus to a small acceptance probability,
while choosing a point close to the initial one will not lead to an efficient exploration of the parameter space, and thus to a slow
convergence of the chain.

Hamiltonian (or "hybrid") dynamics \cite{HMC_Duane1987}, developed originally for lattice field theory, is used to produce candidate proposals for Metropolis algorithm, in a very elegant and efficient
way. It is an exact algorithm which combines molecular dynamics evolution with a Metropolis accept/reject step. This latter is used to correct for discretization errors in the numerical integration of the
corresponding equations of motion. Very good reviews and papers exist, which detail the properties of this algorithm (see for instance \cite{MCMC_Neal}), and only the main ideas will be recalled here for completeness. 

To implement hybrid Monte Carlo algorithm, one introduces for each set of parameters $\hat{q}$ (see previous section) a set of conjugate momenta  $\hat{p}$ and associates to this joint state of "position" $\hat{q}$ and "momentum" $\hat{p}$ an Hamiltonian $H(\hat{q},\hat{p})=\hat{p}^T M^{-1}\hat{p}/2+\mathcal{U}(q)$, where $M$ is a mass matrix, generally taken to be diagonal, and $\mathcal{U}(q)$ an arbitrary potential energy. 
This allows to define a joint distribution as
$$
P(q,p)=\frac{1}{Z}e^{-H(q,p)}=\frac{1}{Z}e^{-\mathcal{K}(p)}e^{-\mathcal{U}(q)}
$$
where $Z$ is the normalizing constant. We use for the potential energy $\mathcal{U}(q)=-\log[P(D|\hat{q})P(\hat{q})]$. 
%The momentum $\hat{p}$ is used as an auxiliary variable, whose 
Starting from a point $\hat{q}_0$ of the chain, the HMC procedure consists in selecting some initial momenta $\hat{p}_0$ normally distributed around zero and let the system evolve deterministically
for some time according to Hamilton's equations of motion for $H(\hat{q},\hat{p})$. It reaches a candidate point $(\hat{q}_1,\hat{p}_1)$ which, 
according to Metropolis procedure described above, is accepted with probability $min(1,e^{-\Delta H})$.
Since the dynamics conserves energy, i.e. $\Delta H=0$ along a trajectory, the acceptance rate is 100\%, independently of the dimension of the vector $\hat{q}$.

%It can be shown that HMC satisfies several properties which are crucial in constructing Markov Chain Monte Carlo updates. 
%Namely, HMC satisfies the detailed balance condition, hence insuring that the points generated with this algorithm
%correctly represent the intended probability distribution, it is reversible, it preserves volume in the phase space (it is symplectic), and it keeps the Hamiltonian invariant.  This latter property implies that the theoretical acceptance probability for Metropolis updates found by Hamiltonian dynamics, 
%s 100\%. 
In practice, this acceptance is degraded because of the numerical resolution of Hamilton equations, but remains still
at a very high level (typically of the order of 70-90\%, independently of the dimension of the chain). 
HMC algorithm is thus very well suited to multi-parameter determination. To discretize Hamilton's equation, we use the LeapFrog
method, a convenient second order integration method that gives the time reversal invariance needed for the Metropolis transition kernel.

We have implemented both Metropolis and Hybrid Hamiltonian Monte Carlo algorithms in the open-source package HeraFitter and its successor xFitter \cite{HeraFitter}. This software
provides a  modular framework to determine PDFs by fitting a large ensemble of experimental data. In what follows, we focus on the proton PDFs
and we use for the PDFs parametrization, the HERAPDF functional 
form, that we just recall here for the sake of clarity: the parametrized HERA PDFs are the valence distribution $xu_v$ and $xd_v$, the gluon
distribution $xg$, and the $\overline{U}$ and $\overline{D}$ distribution defined as 
$x\overline{U}=x\overline{u}$, $x\overline{D}=x\overline{d}+x\overline{s}$. Their functional form
reads
\begin{eqnarray}\label{Eq:funcform}
xf_a(x)=A_a\,x^{B_a}\,(1-x)^{C_a}\,(1+D_ax+E_ax^2)
\end{eqnarray}
where $a$ labels a parton ($g$, $u_{val}$, $d_{val}$, \ldots. See \cite{HeraFitter} for more details). 
The analysis procedure we apply to the Markov Chain we have produced is explained in the next section.

%%%%%%%% Results %%%%%%%%%%%%%%%%%
\section{Markov chain analysis}\label{sec:Analysis}

Assessing statistical errors for observables in Monte Carlo simulations is a subtle task and requires a careful treatment of the Markov
chain. This section presents the different stages of analysis and the checks we have performed.

\bigskip  

The procedure to analyze a Markov Chain consists of several steps. In particular, it is necessary to remove the thermalization (or burn-in)
region, to verify the convergence of the chain and to properly examine correlations between neighboring points in the chain. We have also checked the chain reversibility and the fact that the distribution was correctly sampled. 

%Details of the analysis and first results are explained in the next sections.

%Typical Metropolis chains are shown on Figure \ref{Metro_conv}, where are illustrated the main characteristics of a Markov chain. For illustration and test purposes, we have fixed all parameters to their value given by the standard minimization procedure of xFitter, except parameter $B_g$, which corresponds to XXXXX, and is obtained by a Monte Carlo procedure.   WE SHOULD REMIND THE FUNCTIONAL FORM WE USE.

\subsection{Thermalization}

{\bf The thermalization time} (or {\bf burn-in length}) $b$ of a Markov chain corresponds to a number of states 
$\{\hat{q}_t\}_{t=1,\ldots,b}$ to be discarded from the beginning 
so that the chain forgets its starting point. It can be estimated as being the first state of the random walk -- that is the first 
set of parameters $\hat{q}$, denoted $\hat{q}_{b}$ -- reaching the median value of its target distribution $P_{1/2}$ computed using the entire chain, i.e. 
\begin{equation}\label{P1s2}
 P(\hat{q}_{b}|D) > P_{1/2}
% \pi(\Theta_b) > \pi_{1/2}
\end{equation}

%\color{red} This procedure gives a burn-in length for each parameter under consideration, and we remove from the chain the
%maximum of these various lengths\color{black}.

\begin{figure}[H]
%\begin{center}
%\vspace{-2.3cm}
%\epsfxsize=9cm\epsffile{Graph_HMC_ZMVFNS.eps}
 \center
\includegraphics[width=9cm]{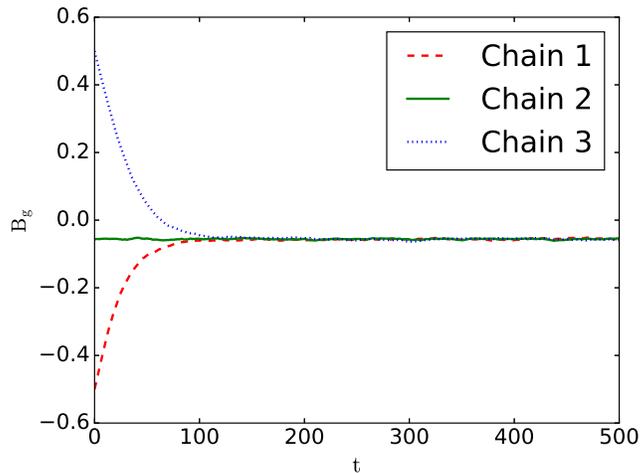}
\caption{{\it {Values of the parameter $B_g$ as a function of the Monte Carlo time for three independent Markov chains. The
green solid line represents a chain starting from the value given by MINUIT minimization, whereas the two other chains start
from values much higher (blue dotted) or much lower (red dashed). The initial corresponding $\chi^2/d.o.f.$ values are, from
chain 1 to chain 3, respectively  $\chi^2/d.o.f.=67.44$, $\chi^2/d.o.f.=0.87$ and $\chi^2/d.o.f.=81.58$. 
We identify clearly on this plot the thermalization region, which is limited to the first $\sim$ 100 - 210 iterations.}}}
\label{Metro_conv}
%\end{center}
\end{figure}

To illustrate thermalization features, we have represented in Figure \ref{Metro_conv} the Monte Carlo  
history of parameter $B_g$ for three independent chains, each starting from a different point. 
For illustration purposes, we have fixed all parameters to their value given by the standard minimization procedure of xFitter, except this parameter $B_g$, which is obtained by a Monte Carlo procedure. 

%These results are obtained from a 10-dimensional chain, and only parameter $B_g$ is plotted here since
%its thermalization features are representative of what happens with all other parameters. 
 
%For illustration, 
%we plot in Figure \ref{Metro_conv} the Monte Carlo history of parameter $B_g$, which corresponds to XXXXX, 
%for three chains starting from different points. These results are obtained from a 10-dimensional Markov Chain. 
%For illustration and test purposes, we have fixed all parameters to their value given by the standard minimization procedure of xFitter, except this parameter $B_g$, which corresponds to XXXXX, and is obtained by a Monte Carlo procedure. 

We can identify in Figure \ref{Metro_conv} the thermalization region, whose extent depends on the starting point. The chain represented in solid green line has been started from the output value of a MINUIT minimization of the $\chi^2$ with respect to parameter $B_g$ and is thus thermalized very quickly, specifically after one iteration, whereas the other chains, started far from the minimum $\chi^2$,
exhibits a thermalization of about 150 iterations for the chain represented in dashed red line and 210 iterations for the one represented by the dotted blue line; as expected, the farther from the minimum the starting point is, the longer the thermalization. 
Starting from a point far from its value given by the minimization procedure is useful to check that this latter did not get stuck in a local minimum and that simulations starting from different points converge to the same region (see also section \ref{subsec:conv}). 
In practice, we have generated several chains (36 chains, to be more precise), starting from random points. We determined the thermalization time using equation (\ref{P1s2}) for each chain, and we removed from each chain its 
burn-in length.

\subsection{Treatment of autocorrelations}

By construction of a Markov chain, the state $\hat{q}_{t}$ depends strongly on the state $\hat{q}_{t-1}$ and quantities computed from this chain form themselves a Markov chain with inherent correlations from one member to the next. 
These type of correlations are often referred to as "autocorrelations" in simulation time.

Let us consider that we would like to extract an observable $O$ from a Markov chain simulation with $N$ points. 
For this estimation, we use the $N$ successive Monte Carlo estimates $O_t$ (we assume in what follows, that the thermalization 
region has already been discarded, i.e. that the chain has been equilibrated before recording data) and we compute the usual
mean $\langle O\rangle$ where $\langle .\rangle$ means averaging over the $N$ data points. 
The usual estimate of root-mean-square deviation of this average can be computed as 
\begin{eqnarray*}
\sigma^2_{naive}=\frac{N}{N-1}\left(\langle O^2\rangle-\langle O\rangle^2\right)
\end{eqnarray*}
This "naive" error relies on the assumption that the measurements performed on the Markov chain are not correlated, which is in general
not true. In order to account for the correlations, one can introduce for the given observable $O$, the integrated autocorrelation time $\tau_{int}$, which can be defined as follows:
\begin{eqnarray*}
\tau_{int}=\frac{1}{2}\displaystyle{\sum_{-\infty}^{+\infty}}\rho(s)
\end{eqnarray*}
where $\rho(s)$ is the normalized autocorrelation function
\begin{eqnarray*}
\rho(s)=\frac{\left(O_t-\langle O\rangle\right)\left(O_{t+s}-\langle O\rangle\right)}{\left(O_{t}-\langle O\rangle\right)^2}
\end{eqnarray*}
The dependence of $\rho(s)$ on the time separation $s$ only is a consequence of the chain being in equilibrium. 
The integrated autocorrelation time controls the statistical error in Monte Carlo measurement of $\langle O\rangle$ and there are mostly two possibilities to incorporate this autocorrelation time in the assessment of the statistical errors. The first one
consists in leaving out $2\tau_{int}$ points between two effective points, or in other terms, to do a subsampling by rejecting all states which are closer than $2\tau_{int}$ to each other, in order to get independent states. This approach has the disadvantage of requiring the
{\it{a priori}} knowledge of $\tau_{int}$. 
The second approach consists in keeping all measurements, but taking into account the autocorrelation time to estimate the statistical
errors. The statistical error of correlated measurements can indeed by computed by \cite{Sokal_1989,Wolff_2003} 
\begin{eqnarray*}
\sigma^2_{\tau}=2\tau_{int}\sigma^2_{naive}
\end{eqnarray*}
This means that the number of "effectively independent data points" in a run of length $N$ is roughly $N/(2\tau_{int})$. 
If the integrated autocorrelation time is used to assess statistical errors, this means of course that a reliable estimate of $\tau_{int}$
and its error itself are needed. Such estimates require a delicate procedure. An efficient method -- called $\Gamma$-method -- has been developed in \cite{Wolff_2003}, which relies on the explicit determination of autocorrelation functions and autocorrelation times. This method
provides not only numerical estimators of the integrated autocorrelation time, but also estimates for mean values and statistical errors
for arbitrary functions of elementary observables in Monte Carlo simulations. We refer the interested reader to \cite{Wolff_2003} and references therein for details. We have used the $\Gamma$-method both to obtain the autocorrelation time and to compute observables.

A further method to reliably estimate the error on uncorrelated measurements is the so called {\it{Jackknife binning}} \cite{Quenouille_1956}. It consists in building $N$ subsets of data from the initial ensemble of size $N$, by removing one observation, leaving samples of size $N-1$.
Pre-averaging over the blocks of data provides $N$ estimates of the average:
\begin{eqnarray*}
%\langle O\rangle_B=\frac{1}{N-J}\left(\displaystyle{\sum_{i=1}^{(B-1)J}}O_i+\displaystyle{\sum_{i=BJ+1}^{N}}O_i\right), \qquad B=1,\ldots, N_{bin}
\langle O\rangle_B= \frac{1}{N-1}\left(\displaystyle{\sum_{t=1}^{(B-1)}}O_t+\displaystyle{\sum_{t=B+1}^{N}}O_t\right), \qquad B=1,\ldots, N
\end{eqnarray*}

The jackknife mean and variance for the observable $O$ are then constructed from:
\begin{eqnarray*}
{\langle O\rangle}_{Jack.}=\frac{1}{N}\displaystyle{\sum_{B=1}^{N}}\langle O\rangle_B,\ \quad
\sigma^2_{Jack.}=\frac{N-1}{N}\displaystyle{\sum_{B=1}^{N}}(\langle O\rangle_B-{\langle O\rangle}_{Jack.})^2
\end{eqnarray*}

The jackknife method -- and its extensions -- is a widely used procedure, in particular in Lattice QCD. For cross-checks and comparison, we applied in our analysis both $\Gamma$-method and jackknife binning technics. For this latter, to un-correlate the points of a given chain, we performed a subsampling of this chain using the value of the autocorrelation time provided by the $\Gamma$-method. 
\subsection{Reversibility and convergence}\label{subsec:conv}

\begin{figure}[H]
%\begin{center}
%\vspace{-1.cm}
\center
\includegraphics[width=9cm,height=6.7cm]{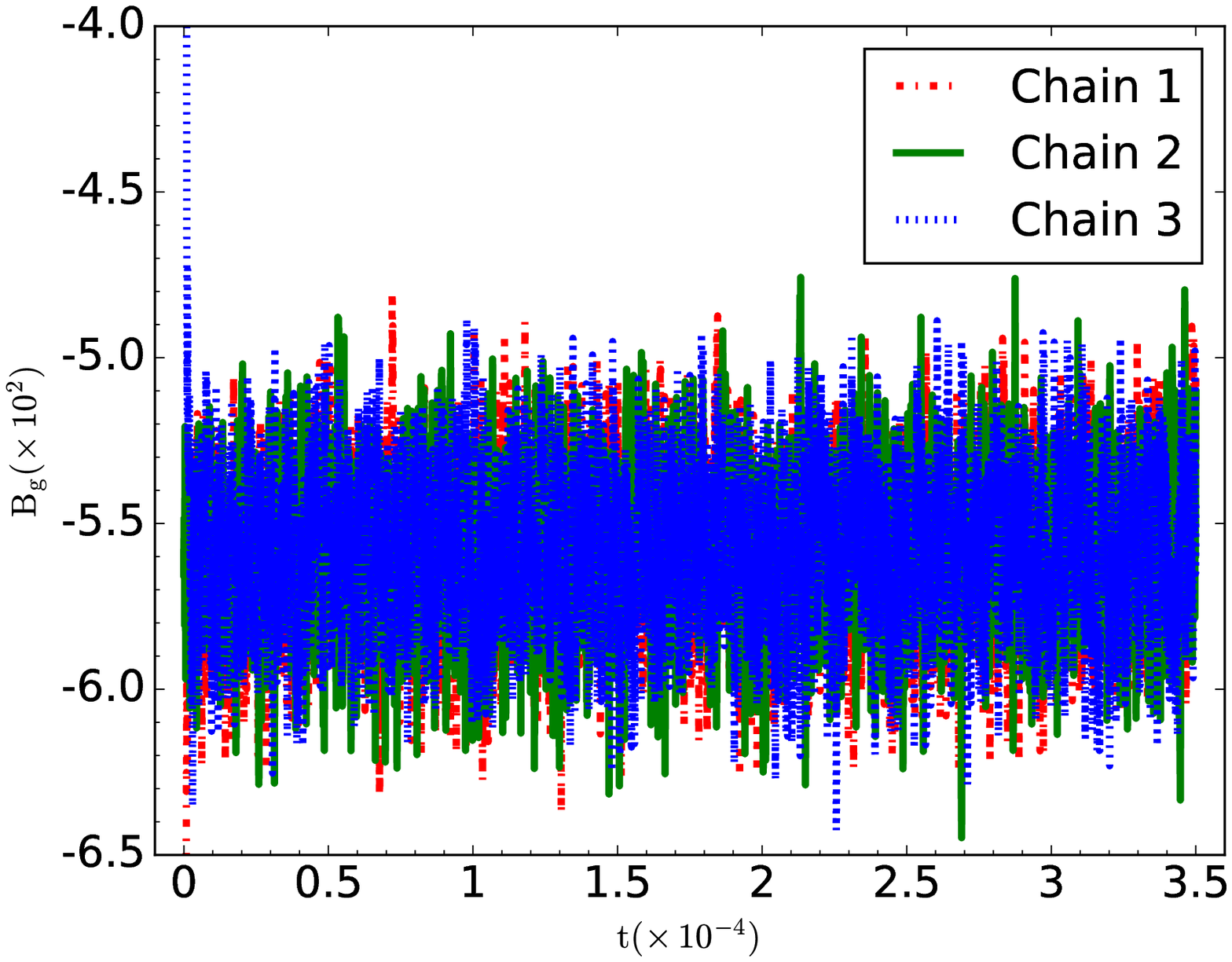}\hspace{-0.5cm}
\includegraphics[width=9cm,,height=6.7cm]{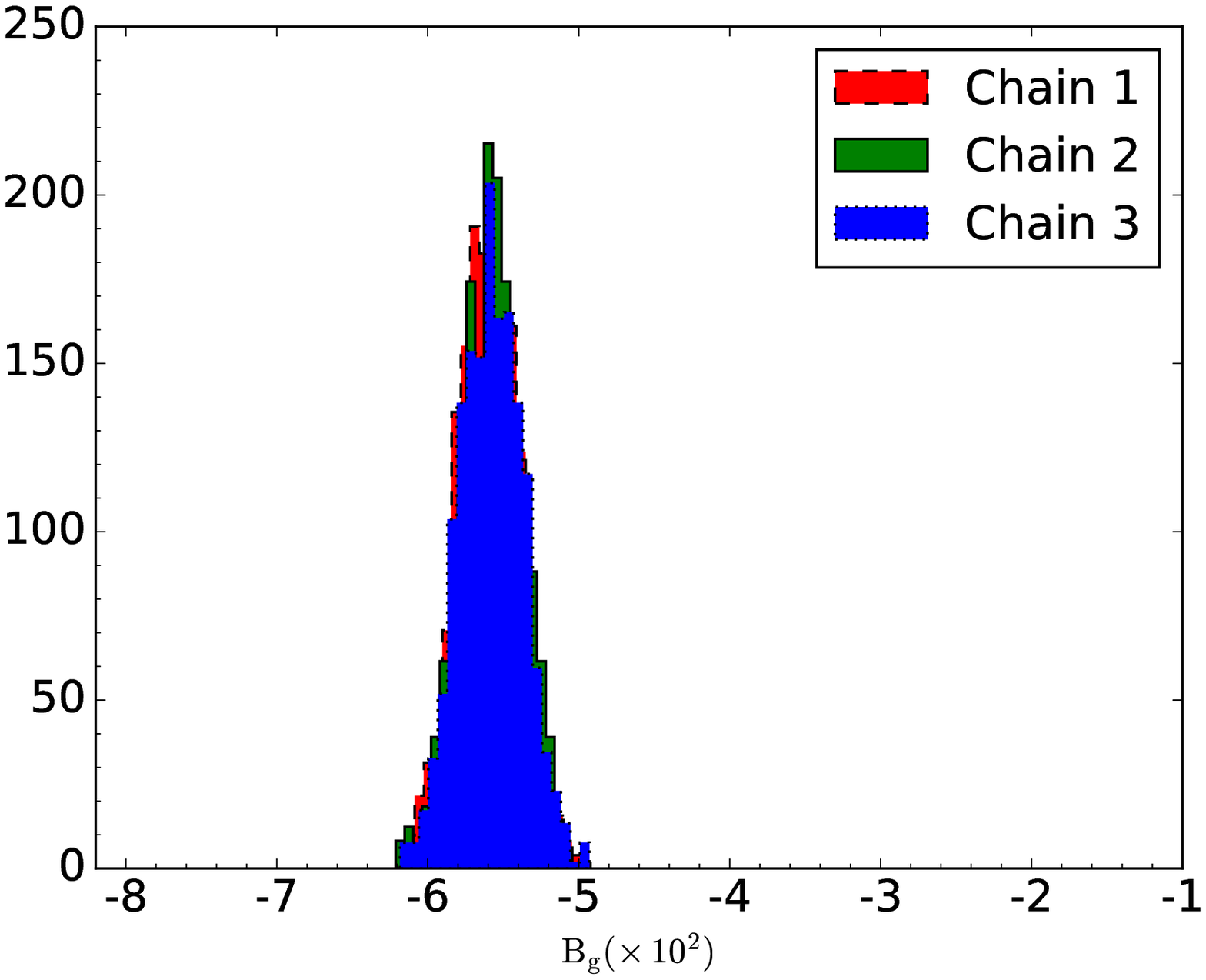}
\caption{{\it {Values of the parameter $B_g$ as a function of the Monte Carlo time for three independent Markov chains (l.h.s.). The starting points of the chains and the color code are the same as in Figure \ref{Metro_conv}. The chains are clearly converging towards the same stationary distribution, what is confirmed by plotting the parameter distribution for each chain (r.h.s., after removing thermalization points and taking into account autocorrelation).    
}}}
\label{HMC_conv}
%\end{center}
\end{figure}

We have verified that our implementation of HMC algorithm satisfies reversibility with a very good precision (relative accuracy better than $10^{-6}$) and that 
 the average acceptance -- computed using jackknife method, after removing thermalization region and decorrelating the chain -- is 
$\langle e^{-\Delta H}\rangle=1.002\pm0.016$, thus insuring that our chains indeed converge towards a stationary distribution.  
In addition, to exclude the risk of a non-identified lack of convergence, we have simulated several chains, with different (and random) 
starting points. This is illustrated in Figure \ref{HMC_conv} in the case of one varying parameter, namely $B_g$ (all others being fixed to their value obtained by standard $\chi^2$ minimization procedure), where the chains are clearly seen to converge towards the same stationary distribution. 

\bigskip

The results displayed in the next section have been obtained after skipping thermalization and properly taking into account the autocorrelation, using either the $\Gamma$-method or the Jackknife binning procedure as explained above.  

%%%%%%%%%%%%%%%%%%%%%%%%%%%%%%%%%%%%%%%%%%%%%%%%%%%%%%%%%%%%%%%%
\section{Preliminary results}\label{sec:Results}

%In this section we show the parton distribution functions obtained from a Markov Chain with 10 parameters, and the settings used to
%produce HERAPDF1.0 distributions, in such a way that we can compare our PDFs with these latter ones. 
\subsection{Set-up and simulation parameters}

The results shown in this section are obtained from a Markov Chain using the HERAPDF
functional forms for initial PDFs at a scale $\mathrm{Q}_0=1.9$ GeV$^2$ with 10 free parameters: $B_g$, $C_g$, $B_{u_{val}}$, $C_{u_{val}}$, 
$E_{u_{val}}$, $C_{d_{val}}$, $C_{\overline{U}}$, $A_{\overline{D}}$, $B_{\overline{D}}$ and $C_{\overline{D}}$ (see expression (\ref{Eq:funcform}) for the definition of these parameters, and \cite{HeraFitter} for more details). We have
used uniform priors for the parameters, and we consider the same data ensembles than the ones used to produce HERAPDF1.0 distributions. 

These data are a combination of inclusive deep inelastic scattering cross-sections measured by the H1 and ZEUS Collaborations in
neutral and charged current unpolarized $e^{\pm}p$ scattering at HERA, during the period 1994-2000 \cite{HeraFitter}. Other settings and cuts 
-- with the exception of the heavy flavor scheme, see below -- were also identical to the ones of HERAPDF1.0 distributions. In particular, we do not rely of K factors, nor on grids techniques. The latter could however interestingly 
be used to speed up the computation\footnote{and we are exploring this possibility for more complete computations.}. 

Since the aim of this work is to demonstrate the applicability of MCMC methods to PDFs determination -- rather than producing competitive PDF sets -- we apply the ZMVFN scheme in order to speed up the computation of the $\chi^2$, and a lower cut on the data at $\mathrm{Q}_{min}^2 = 10$ GeV$^2$. These settings give a total number of data points of 537. We will denote by "HERAPDF1.0 ZMVFNS" the PDFs thus obtained by minimization.

As already mentioned in the previous section, we have generated 36 Monte Carlo chains, each chain 
starting from a different random point, using the HMC algorithm.

The HMC algorithm requires the tuning of essentially two parameters: the number of leapfrog steps $L$ and the step size $\varepsilon$, this latter
potentially depending on the direction in the parameter space. These two quantities are chosen such as to keep both the acceptance high (requiring small $\varepsilon$, to minimize the numerical errors in solving Hamilton equations), and the correlation between two successive Monte Carlo iterations small (thus requiring large trajectory length $L\varepsilon$). We have chosen
$L=100$ and one leapfrog step size for each parameter, depending on the parameter typical standard deviation. Namely we took $\varepsilon_i=3.10^{-2}\Delta q^{typ}_i$, where $\Delta q^{typ}_i$ is, for each parameter, the value of its standard deviation provided by the minimization. 
With these HMC parameters, we obtain an acceptance of 80\% and
chains which have almost no correlation, since the integrated autocorrelation time $\tau_{int}$ computed by the $\Gamma$-method is less than
2 for all parameters. 

The HMC algorithm also requires the computation of the potential energy (that is in our case the $\chi^2$) with respect to the parameters. 
These derivatives are computed numerically, using a symmetric derivative. 
We thus need, for 10 parameters, 20 evaluations of the $\chi^2$ for each step of the Leapfrog algorithm. 
We have run 36 jobs in parallel, 
and collected the results after three days of running. 
We computed for each of our 36 chains the burn-in length, and we removed the maximum burn-in (namely 28) to all of them,
to obtain a total of 4\,400 points per chain. To analyze these chains with the $\Gamma-$method, we kept all
158\,400 points, while the jackknife analysis was done considering one point every four (i.e. $2\tau_{int}$), that is 39\,600 points
\footnote{For a fully realistic PDFs determination, systematic uncertainties (factorization and renormalization scales, heavy quark treatment, \ldots)
are much bigger than the sub-percent accuracy we obtain with such a long Markov Chain. In realistic cases, a Markov Chain with about
a thousand decorrelated points will lead to results statistically accurate enough.}.

\subsection{PDFs parameter values, marginal distributions and correlations}

Table \ref{Tab:GammavsJ} compares the mean value and its statistical error for each of the 10 PDFs parameters under consideration,
using the two analysis procedures we have presented above, namely $\Gamma$-method and jackknife binning. The computation of the integrated autocorrelation time by the $\Gamma$-method gives values of $\tau_{int}$ less than 2 and we have used for the jackknife one point of the chain every four.
As can be seen from Table \ref{Tab:GammavsJ}, both methods give very close results, showing that we have analyze our Markov chain
 in a consistent way. For the rest of this paper, we will thus display only
the results obtained using jackknife binning technics. 
We also notice that for the chain length considered ($\sim$ 40 000 thermalized and decorrelated points), the statistical errors on the mean values are tiny.

\begin{table}[htbp]\begin{center}
\caption{{\it{Comparison of mean values and their statistical errors obtained for PDFs parameters using two different analysis methods. The jackknife binning has been applied after subsampling the chain, selecting points not closer than $2\tau_{int}=4$ from each other. Since the number of points
considered for the jackknife analysis is thus four times less than for the $\Gamma-$method, the errors are expected to be smaller by roughly a factor 2 for
this latter procedure, which is indeed the case.\\
}}}
\begin{tabular}{|c|c|c|}\hline\label{Tab:GammavsJ}
parameter  & $\Gamma$-method   & jackknife binning \\\hline\hline
 ${\bf B}_{\bf g}$   &   $-0.0537 \pm 0.0001$& $-0.0537\pm 0.0002$ \\
 ${\bf C}_{\bf g}$  & $5.9449\pm0.0015$ &  $5.9483\pm 0.0025$ \\
 ${\bf B}_{\mathrm{\bf u}_{\mathrm{\bf val}}}$  &$0.6124\pm0.0001$  &  $0.6125 \pm 0.0002$ \\
 ${\bf C}_{\mathrm{\bf u}_{\mathrm{\bf val}}}$  & $4.7458\pm0.0003$ &  $4.7455 \pm 0.0006 $ \\ 
 ${\bf E}_{\mathrm{\bf u}_{\mathrm{\bf val}}}$  & $14.965\pm0.008$ &  $14.961 \pm 0.012$ \\ 
 ${\bf C}_{\mathrm{\bf d}_{\mathrm{\bf val}}}$  & $3.2054\pm0.0014$  & $3.2077 \pm 0.0016$  \\ 
${\bf C}_{\overline{\bf U}}$  & $4.0917\pm0.0038$ &  $4.0961 \pm 0.0048$ \\ 
${\bf A}_{\overline{\bf D}}$  & $0.3096\pm0.0002$ &  $0.3098 \pm 0.0002 $ \\ 
${\bf B}_{\overline{\bf D}}$  &  $-0.0174\pm0.0001$ &  $-0.0173 \pm 0.0002$ \\ 
${\bf C}_{\overline{\bf D}}$  & $6.2203\pm0.0054$ &  $6.2096 \pm 0.0076$ \\ \hline
\end{tabular}
\end{center}
\end{table}%

\bigskip

In Table \ref{Tab:MCMCvsMinuit} are displayed the results provided by MCMC method  -- using jackknife binning for error estimate -- for the parameters mean, best fit\footnote{The parameter best fit values are the parameters values that minimized the $\chi^2$ function.} and standard deviation\footnote{we have computed here the corrected sample standard deviation}, 
compared with the output of the standard MINUIT minimization. 
\begin{table}[htbp]
\caption{{\it{Comparison of results obtained for PDFs parameter values extracted from an independent Markov Chain of length 39 600, and the results provided
from a MINUIT minimization. We compute from the MCMC the %mean value
 best fit value, the mean value and the standard deviation value for each of
the 10 parameters considered, together with their statistical errors estimated by jackknife binning method. The standard deviation given by the minimization is the usual one-sigma deviation.}}}
\begin{center}
\begin{tabular}{|l|l|l|l|}
\hline
%parameter \#  &      & MCMC   & MINUIT minimization   \\\hline
%\hline
\multicolumn{1}{|c|}{parameter}& \multicolumn{1}{c|}{values}& \multicolumn{1}{c|}{MCMC} &  \multicolumn{1}{c|}{MINUIT minimization}\\
\hline\hline
 %\multirow{2}{*}{\bf $B_g$}& \multicolumn{1}{c|}{mean}&$\quad-0.0537\pm0.0002\quad$ & \ $\qquad-0.0559\qquad$\\
 \multirow{3}{*}{ ${\bf B}_{\bf g}$}
 & \multicolumn{1}{c|}{mean}& $\quad-0.0537\pm 0.0002\quad$& \\
%\cline{2-4}
& \multicolumn{1}{c|} {best fit} &$\quad-0.0632\pm0.0168\quad$ & \ $\qquad-0.0559\qquad$\\
% \cline{2-4}
%\bf $B_g$& & &  \\
& \multicolumn{1}{c|}{standard deviation}& $\quad0.0299\pm0.0001\quad$& \ $\qquad0.0288\qquad$\\
\hline
%\multirow{2}{*}{\bf $C_g$}& \multicolumn{1}{c|}{mean}&$\quad5.9483\pm0.0025\quad$ & \ $\qquad5.9274\qquad$\\
\multirow{3}{*}{ ${\bf C}_{\bf g}$}
& \multicolumn{1}{c|}{mean}& $\quad 5.9483\pm 0.0025 \quad$& \\
%\cline{2-4}
& \multicolumn{1}{c|}{best fit}&$\quad5.8952\pm0.0615\quad$ & \ $\qquad5.9274\qquad$\\
% \cline{2-4}
%\bf $B_g$& & &  \\
& \multicolumn{1}{c|}{standard deviation}& $\quad0.5037\pm0.0019\quad$&\ $\qquad0.5078\qquad$ \\
\hline
%\multirow{2}{*}{\bf $B_{uval}$}& \multicolumn{1}{c|}{mean}& $\quad0.6125\pm0.0002\quad$&\ $\qquad0.6098\qquad$\\
\multirow{3}{*}{ ${\bf B}_{\mathrm{\bf u}_{\mathrm{\bf val}}}$}
& \multicolumn{1}{c|}{mean}& $\quad 0.6125 \pm 0.0002 \quad$& \\
%\cline{2-4}
& \multicolumn{1}{c|}{best fit}& $\quad0.6092\pm0.0121\quad$&\ $\qquad0.6098\qquad$\\
% \cline{2-4}
%\bf $B_g$& & &  \\
& \multicolumn{1}{c|}{standard deviation}& $\quad0.0371\pm0.0001\quad$&\ $\qquad0.0389\qquad$ \\
\hline
%\multirow{2}{*}{\bf $C_{uval}$}& \multicolumn{1}{c|}{mean}& $\quad4.7455\pm 0.0006\quad$&\ $\qquad4.7122\qquad$\\
\multirow{3}{*}{${\bf C}_{\mathrm{\bf u}_{\mathrm{\bf val}}}$}
& \multicolumn{1}{c|}{mean}& $\quad  4.7455 \pm 0.0006\quad$& \\
%\cline{2-4}
& \multicolumn{1}{c|}{best fit}& $\quad4.7467\pm 0.0525\quad$&\ $\qquad4.7122\qquad$\\
% \cline{2-4}
%\bf $B_g$& & &  \\
& \multicolumn{1}{c|}{standard deviation}& $\quad0.1280\pm0.0005\quad$& \ $\qquad0.1332\qquad$\\
\hline
%\multirow{2}{*}{\bf $E_{uval}$}& \multicolumn{1}{c|}{mean}& $\quad14.9611\pm 0.0125\quad$&\ $\qquad14.7589\qquad$\\
\multirow{3}{*}{ ${\bf E}_{\mathrm{\bf u}_{\mathrm{\bf val}}}$}
& \multicolumn{1}{c|}{mean}& $\quad 14.961 \pm 0.012 \quad$& \\
%\cline{2-4}
& \multicolumn{1}{c|}{best fit}& $\quad15.42\pm 0.94\quad$&\ $\qquad14.76\qquad$\\
% \cline{2-4}
%\bf $B_g$& & &  \\
& \multicolumn{1}{c|}{standard deviation}& $\quad2.494\pm0.010\quad$& \ $\qquad2.571\qquad$\\
\hline
%\multirow{2}{*}{\bf $C_{dval}$}& \multicolumn{1}{c|}{mean}& $\quad3.2077\pm0.0016\quad$ &\ $\qquad3.1435\qquad$\\
\multirow{3}{*}{ ${\bf C}_{\mathrm{\bf d}_{\mathrm{\bf val}}}$}
& \multicolumn{1}{c|}{mean}& $\quad 3.2077 \pm 0.0016 \quad$& \\
%\cline{2-4}
& \multicolumn{1}{c|}{best fit}& $\quad3.084\pm0.076\quad$ &\ $\qquad3.143\qquad$\\
% \cline{2-4}
%\bf $B_g$& & &  \\
& \multicolumn{1}{c|}{standard deviation}& $\quad0.3183\pm0.0016\quad$&\ $\qquad0.2830\qquad$ \\
\hline
%\multirow{2}{*}{\bf $C_{\overline{U}}$}& \multicolumn{1}{c|}{mean}& $\quad4.0961\pm0.0048\quad$&\ $\qquad4.0520\qquad$\\
\multirow{3}{*}{ ${\bf C}_{\overline{\bf U}}$}
& \multicolumn{1}{c|}{mean}& $\quad 4.0961 \pm 0.0048 \quad$& \\
%\cline{2-4}
& \multicolumn{1}{c|}{best fit}& $\quad4.323\pm0.814\quad$&\ $\qquad4.052\qquad$\\
% \cline{2-4}
%\bf $B_g$& & &  \\
& \multicolumn{1}{c|}{standard deviation}& $\quad0.9617\pm0.0035\quad$ &\ $\qquad0.9782\qquad$ \\
\hline
%\multirow{2}{*}{\bf $A_{\overline{D}}$}& \multicolumn{1}{c|}{mean}& $\quad0.3098\pm0.0002\quad$ &\ $\qquad0.3051\qquad$\\
\multirow{3}{*}{${\bf A}_{\overline{\bf D}}$}
& \multicolumn{1}{c|}{mean}& $\quad 0.3098 \pm 0.0002 \quad$& \\
%\cline{2-4}
& \multicolumn{1}{c|}{best fit}& $\quad0.320\pm0.039\quad$ &\ $\qquad0.305\qquad$\\
% \cline{2-4}
%\bf $B_g$& & &  \\
& \multicolumn{1}{c|}{standard deviation}& $\quad0.0485\pm0.0002\quad$ & \ $\qquad0.0488\qquad$\\
\hline
%\multirow{2}{*}{\bf $B_{\overline{D}}$}& \multicolumn{1}{c|}{mean}& $\quad-0.0173\pm0.0002\quad$ &\ $\qquad-0.0178\qquad$\\
\multirow{3}{*}{ ${\bf B}_{\overline{\bf D}}$}
& \multicolumn{1}{c|}{mean}& $\quad -0.0173 \pm 0.0002 \quad$& \\
%\cline{2-4}
& \multicolumn{1}{c|}{best fit}& $\quad-0.0097\pm0.0237\quad$ &\ $\qquad-0.0178\qquad$\\
% \cline{2-4}
%\bf $B_g$& & &  \\
& \multicolumn{1}{c|}{standard deviation}& $\quad0.0304\pm0.0001\quad$ & \ $\qquad0.0306\qquad$\\
\hline
%\multirow{2}{*}{\bf $C_{\overline{D}}$}& \multicolumn{1}{c|}{mean}& $\quad6.2096\pm0.0076\quad$ &\ $\qquad5.8753\qquad$\\
\multirow{3}{*}{ ${\bf C}_{\overline{\bf D}}$}
& \multicolumn{1}{c|}{mean}& $\quad 6.2096 \pm 0.0076 \quad$& \\
%\cline{2-4}
& \multicolumn{1}{c|}{best fit}& $\quad5.888\pm0.142\quad$ &\ $\qquad5.875\qquad$\\
% \cline{2-4}
%\bf $B_g$& & &  \\
& \multicolumn{1}{c|}{standard deviation}& $\quad1.505\pm0.009\quad$  & \ $\qquad1.290\qquad$\\
\hline
%\\\hline              
\end{tabular}
\end{center}
\label{Tab:MCMCvsMinuit}
\end{table}%
We notice, as already stated above, that the determination of parameters by Monte Carlo methods, gives much more information than a minimization. We can in particular extract the statistical errors on the quantities we are interested in and this error decreases with the length of the Markov chain. As seen from the table, the most probable value extracted from the parameter probability distributions are compatible
with the parameter values provided by the minimization procedure. For what concerns
standard deviations however, although MCMC and minimization gives similar results, no precise
comparison can be made since the minimization does not provide estimates of errors for
the usual one-standard deviation of the parameters. In addition, both quantities (MCMC standard
deviation and minimization deviation) should coincide only if the probability density of the
parameter considered is Gaussian, which is not necessarily true (see below). This fact is already
visible in Table \ref{Tab:MCMCvsMinuit}, where we can see that the mean and best fit 
values do not coincide within errors for some of the parameters ($C_{d_{val}}$, $C_{\overline{D}}$
for instance). 

\begin{figure}[H]
\begin{center}
%\vspace{8cm}
%\mbox{\epsfxsize=15cm\epsfysize=7cm\epsffile{hist_HMC_10D_ZMVFNS.eps}}
%\includegraphics[width=18cm]{hist_HMC_10D_ZMVFNS.eps}
\includegraphics[width=18cm]{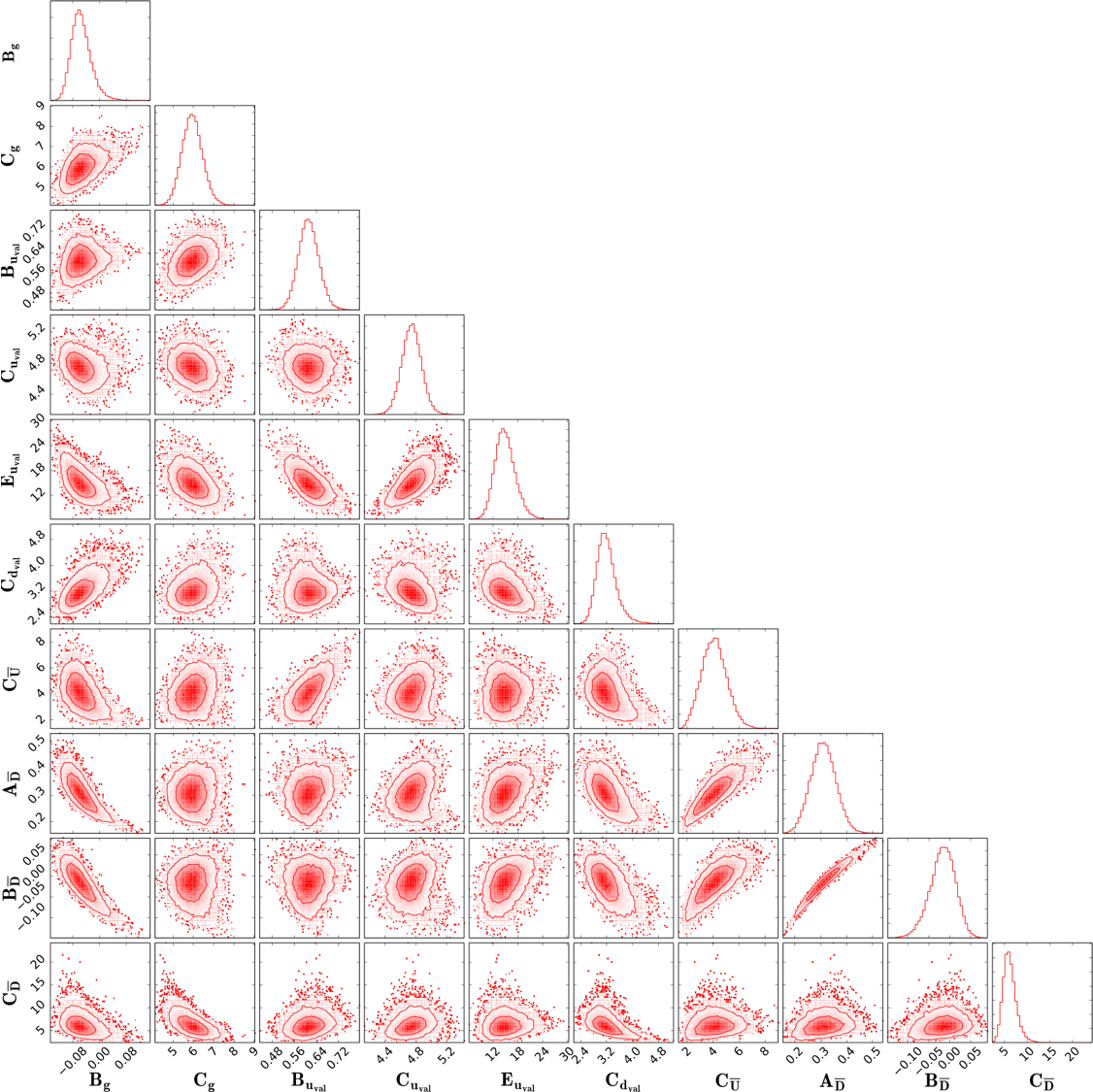}
\end{center}
\vspace{-0.5cm}
\caption{{\it Probability distribution functions of the PDFs parameters (diagonal) and 2D correlation plots between parameters (off-diagonal).}}\label{Fig:paradistr}
\end{figure}

\begin{figure}[H]
\begin{center}
%\mbox{\epsfxsize=8cm\epsfysize=6cm\epsffile{Chi2_HMC_10D_ZMVFNS.eps}}
\includegraphics[width=8.7cm]{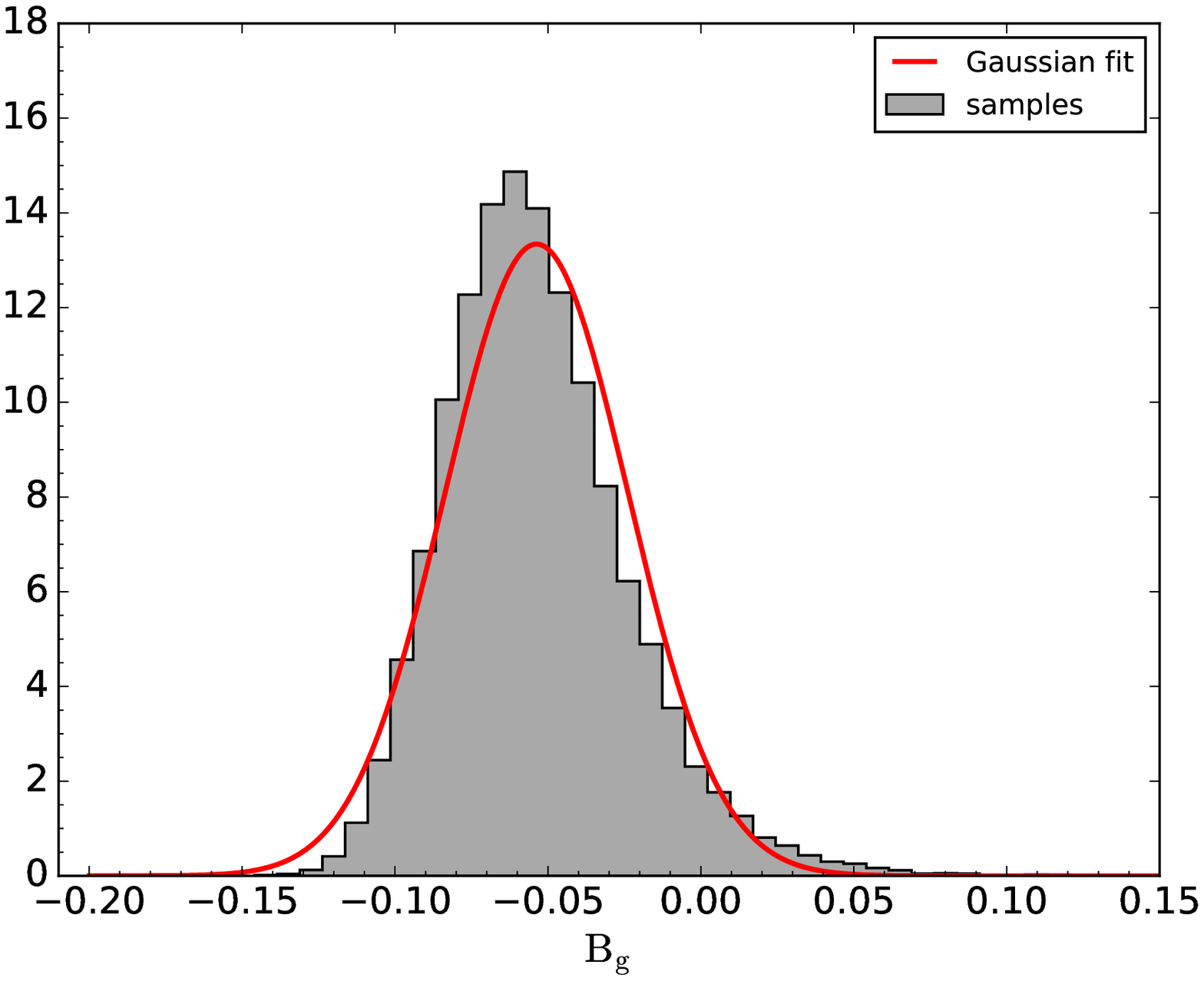}
\includegraphics[width=8.7cm]{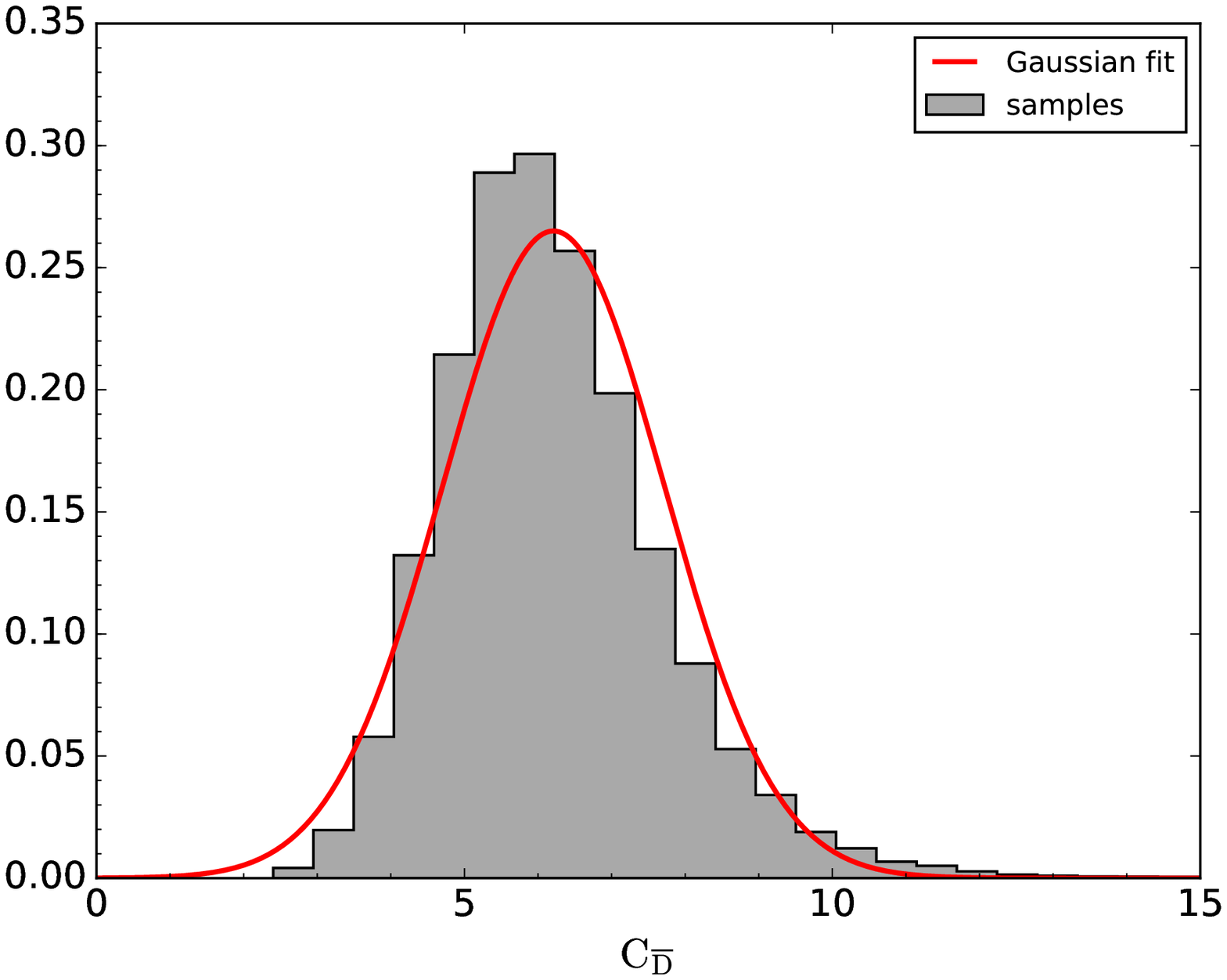}
\end{center}
\vspace{-0.5cm}
\caption{{\it Marginal probability distribution of parameters $B_g$ and $C_{\overline{D}}$.
They do not follow a Gaussian law, as can be seen from the gaussian fit (solid red line).
}}\label{MargDistr}
\end{figure}

%As mentioned in the first section, MCMC provides marginal probability distributions of the PDFs free parameters, and confidence intervals for these parameters. 
The probability distribution functions of the parameters, together with the 2D correlation plots between parameters are displayed in Figure \ref{Fig:paradistr}.  The marginal posterior parameter distributions are shown on the diagonal graphs, and
2D-correlations on the off-diagonal plots.  The inner and outer contours of these latter are taken to be regions containing 
respectively 68\% and 95\% of the probability density. We note that the probability distribution of
some parameters cannot properly be described by a Gaussian law, as illustrated in Figure \ref{MargDistr}, and are even non symmetric.  
We have also checked that our correlation plots are compatible with the values
provided by the covariant matrix.

\begin{figure}[htbp]
\begin{center}
%\mbox{\epsfxsize=8cm\epsfysize=6cm\epsffile{Chi2_HMC_10D_ZMVFNS.eps}}
\includegraphics[width=9cm]{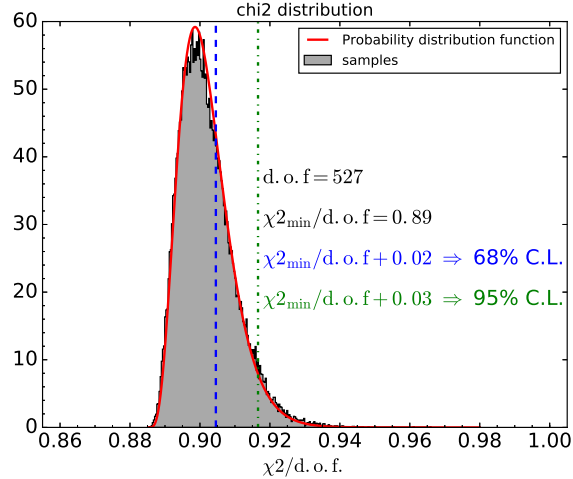}
\end{center}
\vspace{-0.5cm}
\caption{{\it  $\chi^2$ distribution for a 10D Monte Carlo chain. The solid red line is an adjustment
of these data with a $\chi^2$ distribution law with 10 degrees of freedom. 
The dashed (dashdot) vertical line indicates the 68\% (95\%) confidence limit.}}\label{Chi2}
\end{figure}

In Figure \ref{Chi2} is shown the $\chi^2/d.o.f.$ distribution for our 10-dimensional Monte Carlo chain. The solid red line is an adjustment with a $\chi^2$ distribution law with 10 degrees of freedom, 
which perfectly describes our results. 
The fact that the $\chi^2$ function built in equation (\ref{eq:Like}) follows a $\chi^2$ distribution law with the expected number of degrees of freedom is a strong indication (though not a formal proof) that our assumptions concerning the fluctuations of the experimental data points around their corresponding theoretical values, are justified. 

Though this is not the case in the example we present, we would also like to note that potential flat directions in the parameter space 
are less problematic for MCMC method than for minimization techniques. %is not a problem 
%for MCMC method since the goal is not to minimize the $\chi^2$. 
%MCMC technics also leave the possibility to use over-flexible functional forms for the PDFs and therefore less parametrization bias. 
\par
We need now to calculate, from the Markov chain of parameters, the parton distribution
functions we are interested in. The procedure we use is explained in the next section, and
is more generally valid for any observable we want to compute from the MCMC.

\subsection{PDFs marginal distributions and confidence interval}

\begin{figure}[H]
\begin{center}
\includegraphics[width=8.8cm,height=7cm]{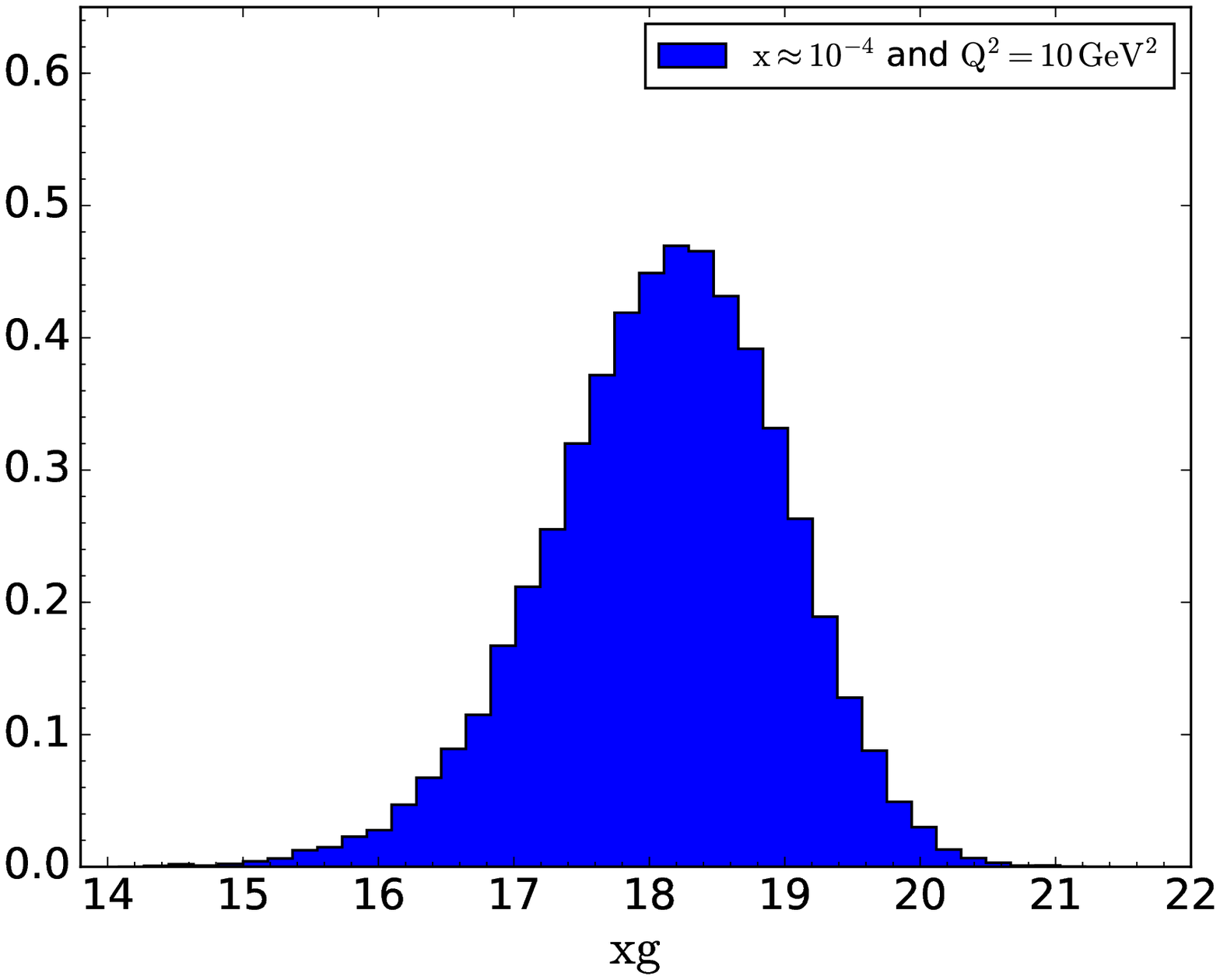}
\includegraphics[width=8.8cm,height=7cm]{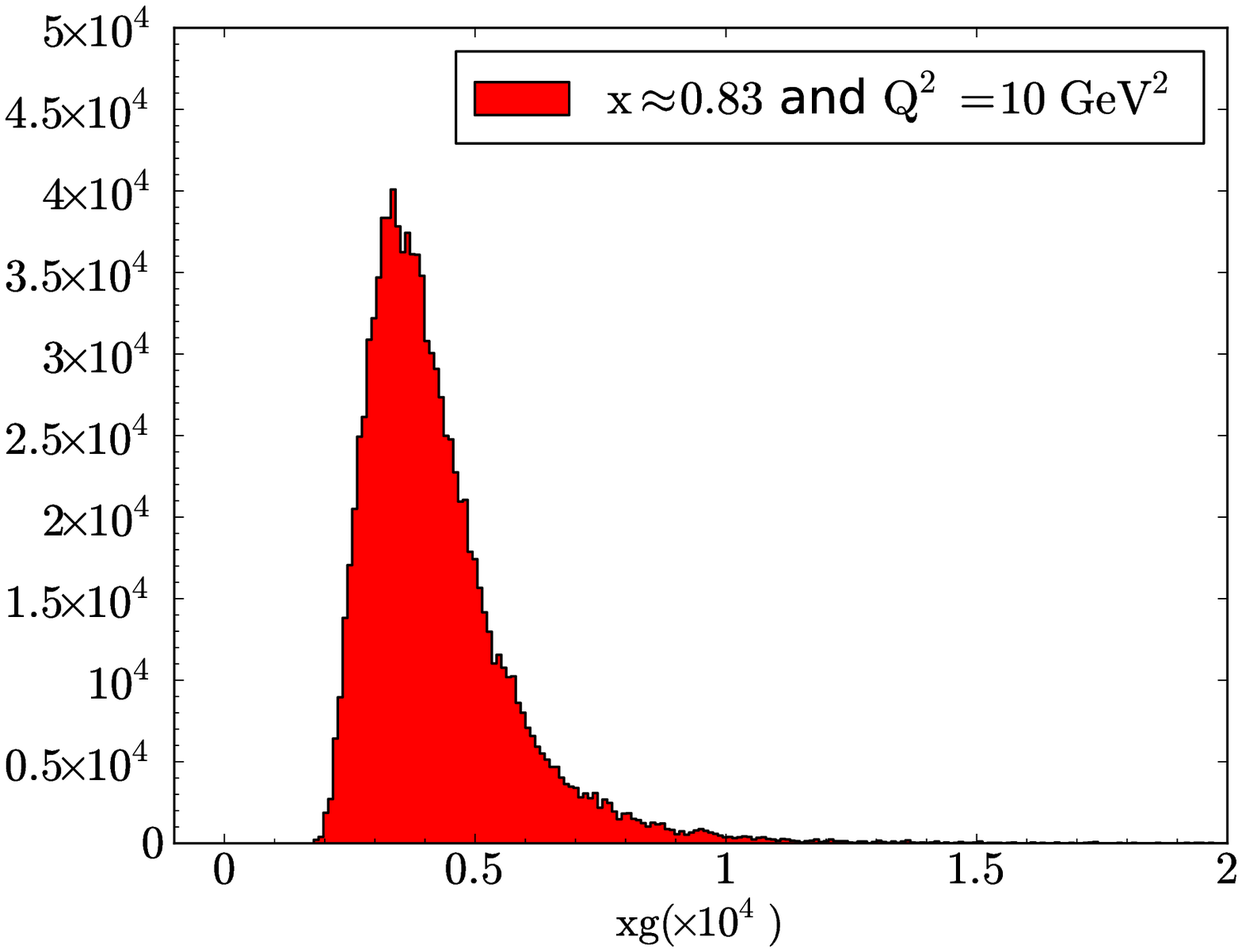}
\end{center}
\vspace{-0.3cm}
\caption{{\it Gluon PDF probability distribution function for $x\approx10^{-4}$ (l.h.s.) and $x\approx 0.83$ at fixed $Q^2=10$ GeV$^2$. The $68\%$
confidence interval is obtained from this distribution, 
considering the region of the distribution containing $68$\% of the data remaining on each side of the best fit value.}}\label{PDF_dist}
\end{figure}

%EXPLAIN HOW WE COMPUTED PDF
%SHOW PDF PDF
%COMMENTS ON CORRELATIONS AND HOW WE CHOOSE CONFIDENCE INTERVALS.
To extract parton distribution functions from the Markov chain, we compute, from 
the set of 10 parameters obtained at each Monte Carlo iteration, the corresponding 
PDFs for a range of $x$ and $Q^2$ values. This provides the marginal probability density
functions of PDFs at fixed $(x,Q^2)$, as illustrated on Figure \ref{PDF_dist} for the gluon, for two different $x$ values.

\begin{figure}[H]
{\includegraphics[width=9cm,height=6.1cm]{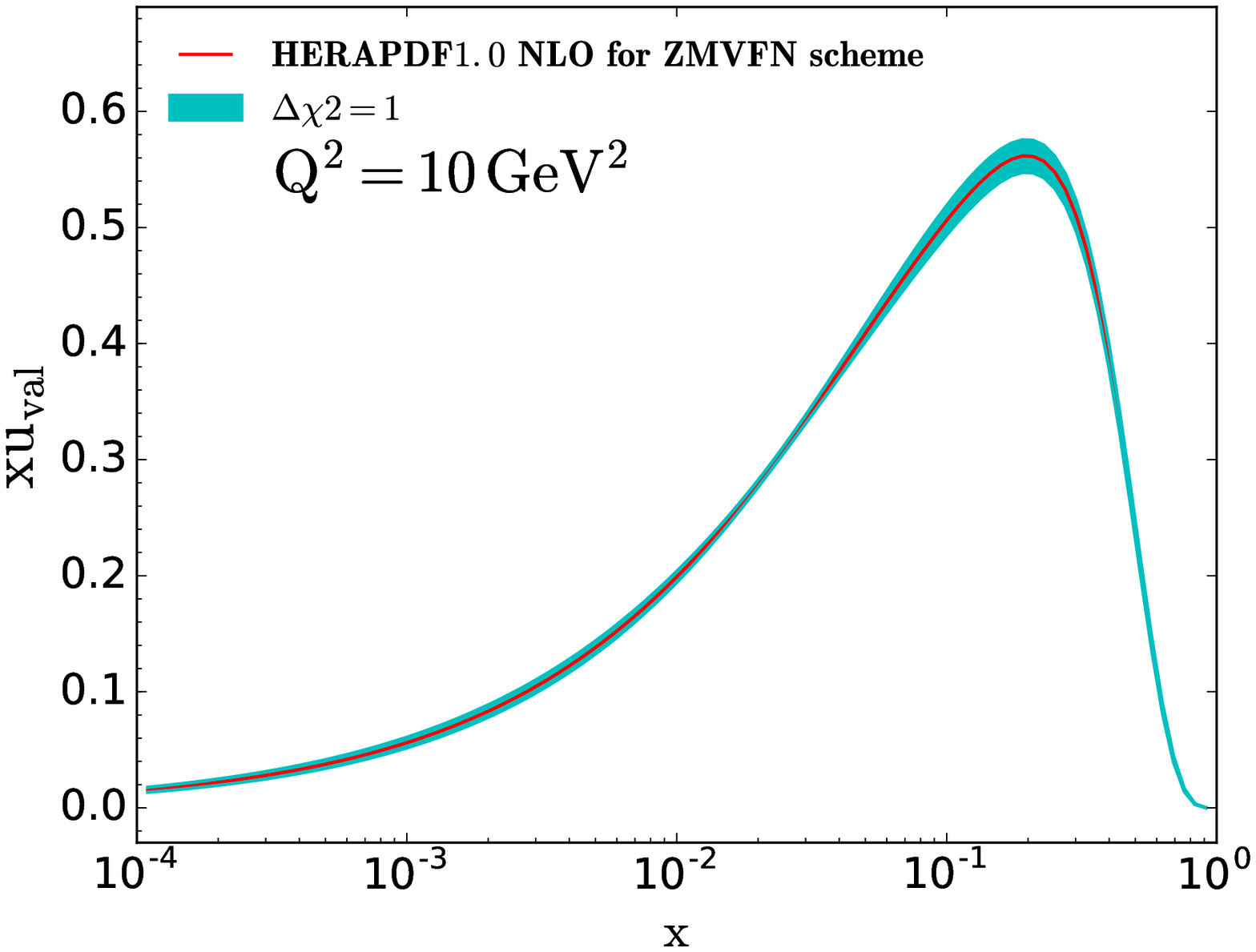}}%\hspace{-0.5cm}
%\raisebox{-0.26mm}
{\includegraphics[width=9cm,height=6.1cm]{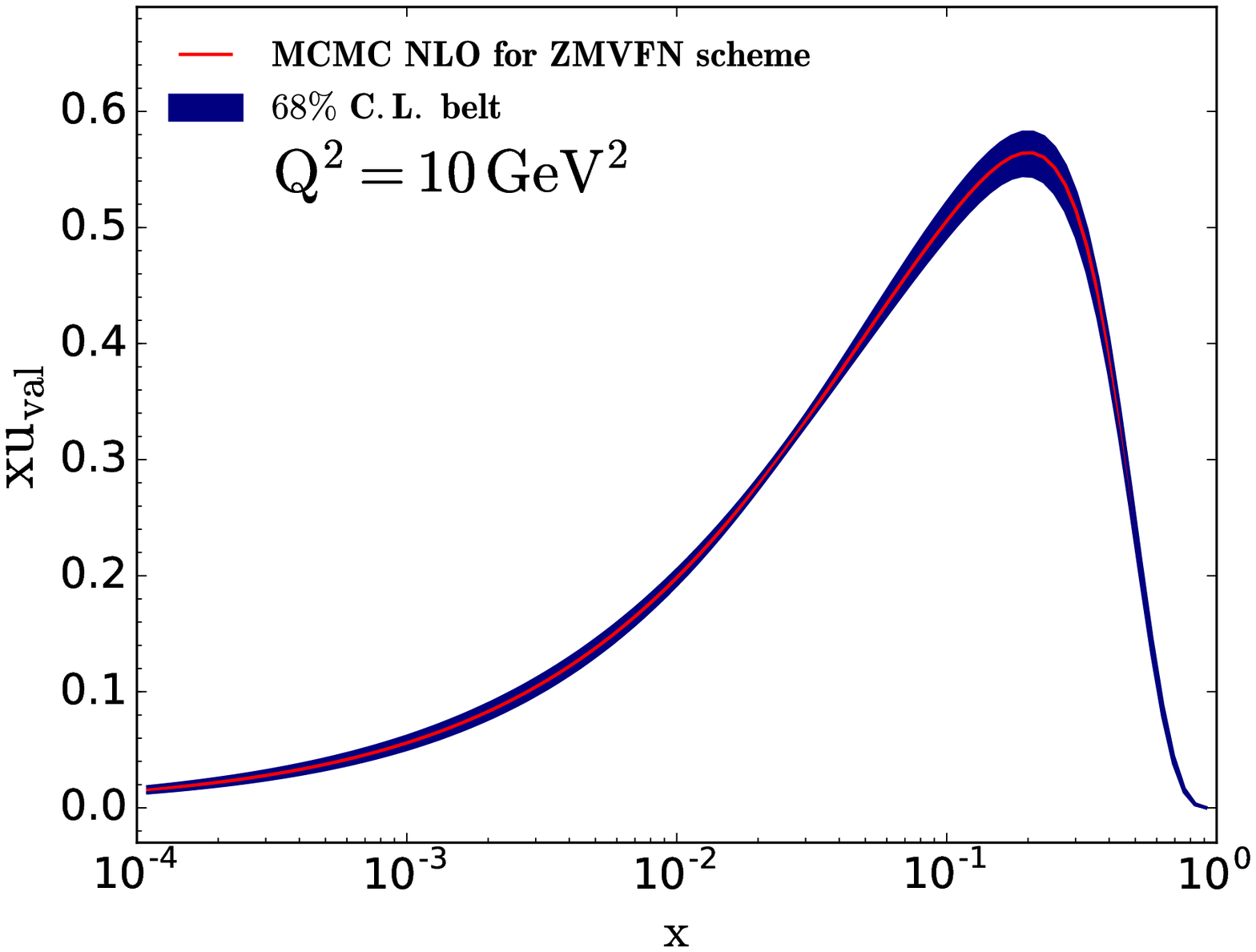}}
\includegraphics[width=9cm,height=6.1cm]{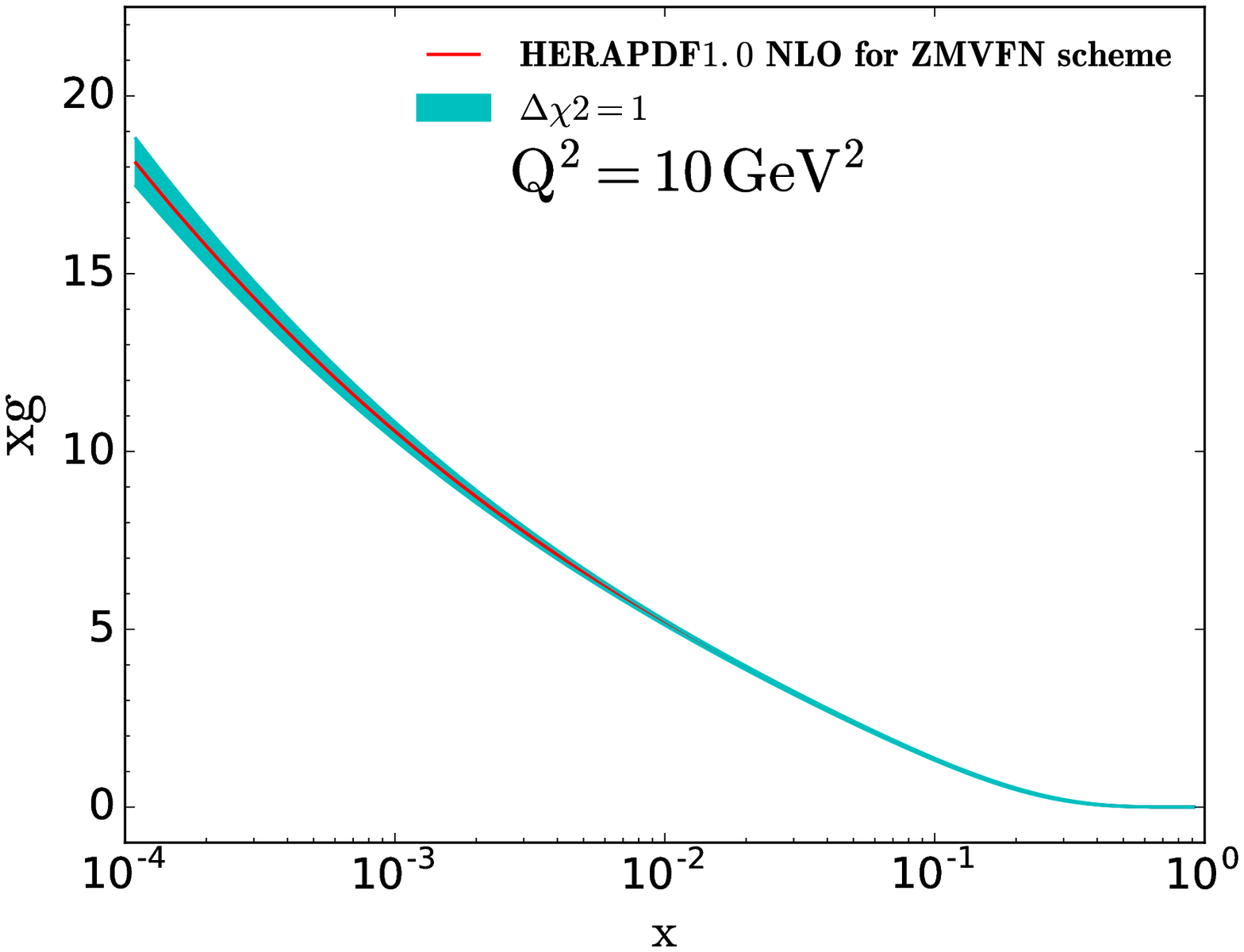}
%\raisebox{-0.26mm}
{\includegraphics[width=9cm,height=6.1cm]{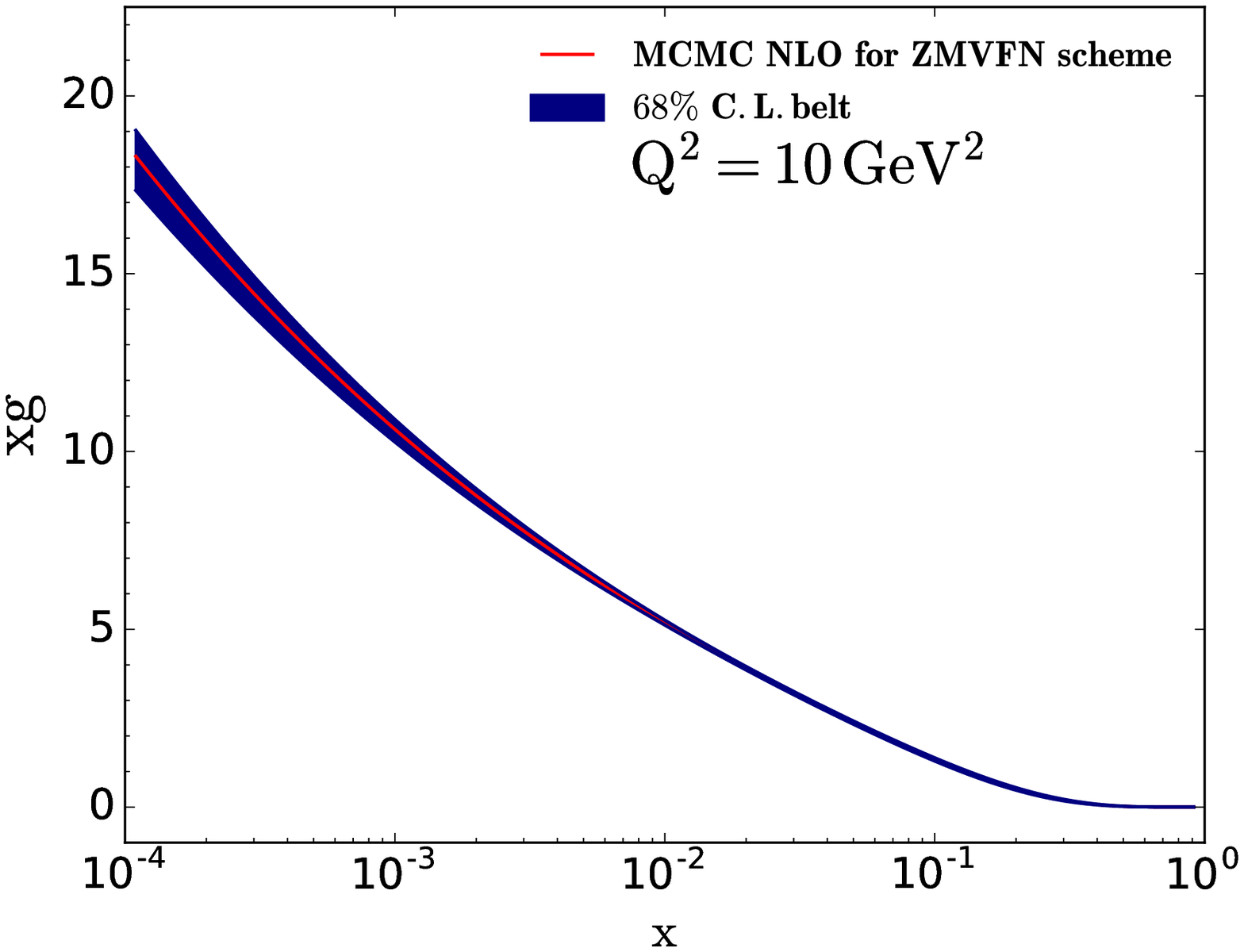}}

\vspace{-0.2cm}
\caption{{\it The parton distribution functions obtained using MCMC (right) compared to HERAPDF1.0 (ZMVFN scheme) from xFitter output (left) for $\mathrm{xu}_{\mathrm{val}}$ and $\mathrm{xg}$,
at $\mathrm{Q^2}$ = $10\ \mathrm{GeV^2}$. The bands show the $68\%$ confidence interval around the central value (in solid red line)
for the MCMC PDFs, and the standard $\Delta\chi^2=1$ deviation for HERAPDF.}}
\label{PDfs_uval_Gluon}
\end{figure}

\begin{figure}[H]
\begin{center}
%\vspace{8cm}
%\mbox{\epsfxsize=15cm\epsfysize=7cm\epsffile{hist_HMC_10D_ZMVFNS.eps}}
%\includegraphics[width=8.9cm]{Gluon_Sea_uval_dval_02_pdfs_SEMILOG_RATIO.eps}%\hspace{-0.6cm}
\includegraphics[width=8.9cm,,height=6.5cm]{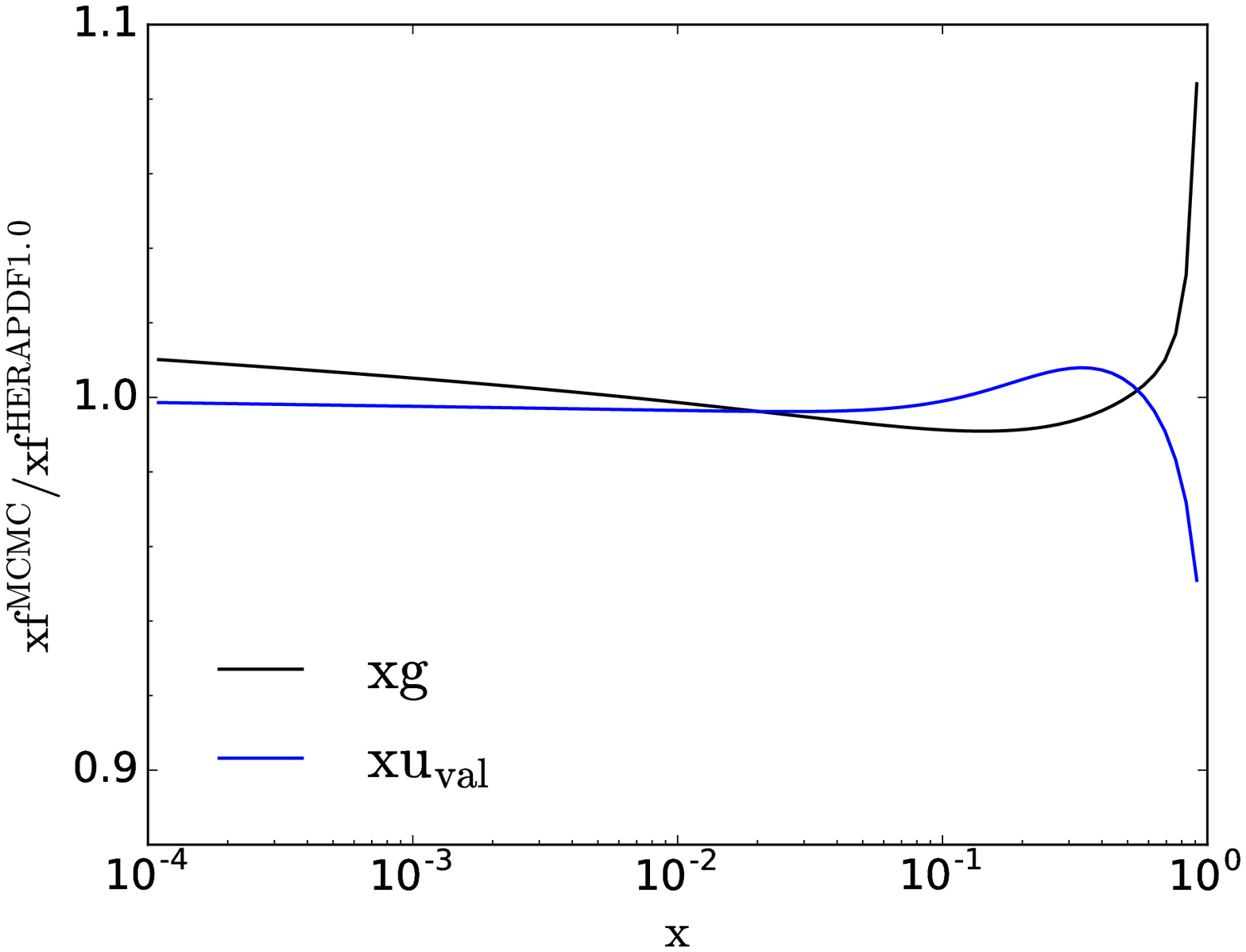}%\hspace{-0.6cm}
\end{center}
\vspace{-0.5cm}
\caption{ {\it{Ratio of MCMC PDFs and HERAPDF1.0 (ZMVFN scheme) central values for $\mathrm{xu}_{\mathrm{val}}$ and $\mathrm{xg}$ at $\mathrm{Q^2}=10~\mathrm{GeV^2}$.}}
}
%\it The parton distribution functions obtained using MCMC4HERAPDF1.0 in the $\mathrm{ZMVFNS\ scheme}$ for $\mathrm{xu}_{\mathrm{val}}$, $\mathrm{xd}_{\mathrm{val}}$, $\mathrm{x\Sigma} = \mathrm{x\overline{u}} + \mathrm{x\overline{d}} + \mathrm{xs} + \mathrm{xc}$\ and\ $\mathrm{xg}$,
%at $\mathrm{Q^2}$ = $10\ \mathrm{GeV^2}$. The bands show
%experimental uncertainties ($68\%$ confidence limits) of the MCMC fit.}}
\label{Fig:Ratio_Gluon_Sea_uval_dval_pdfs}
\end{figure}

\begin{figure}[H]
{\includegraphics[width=9cm,height=7.0cm]{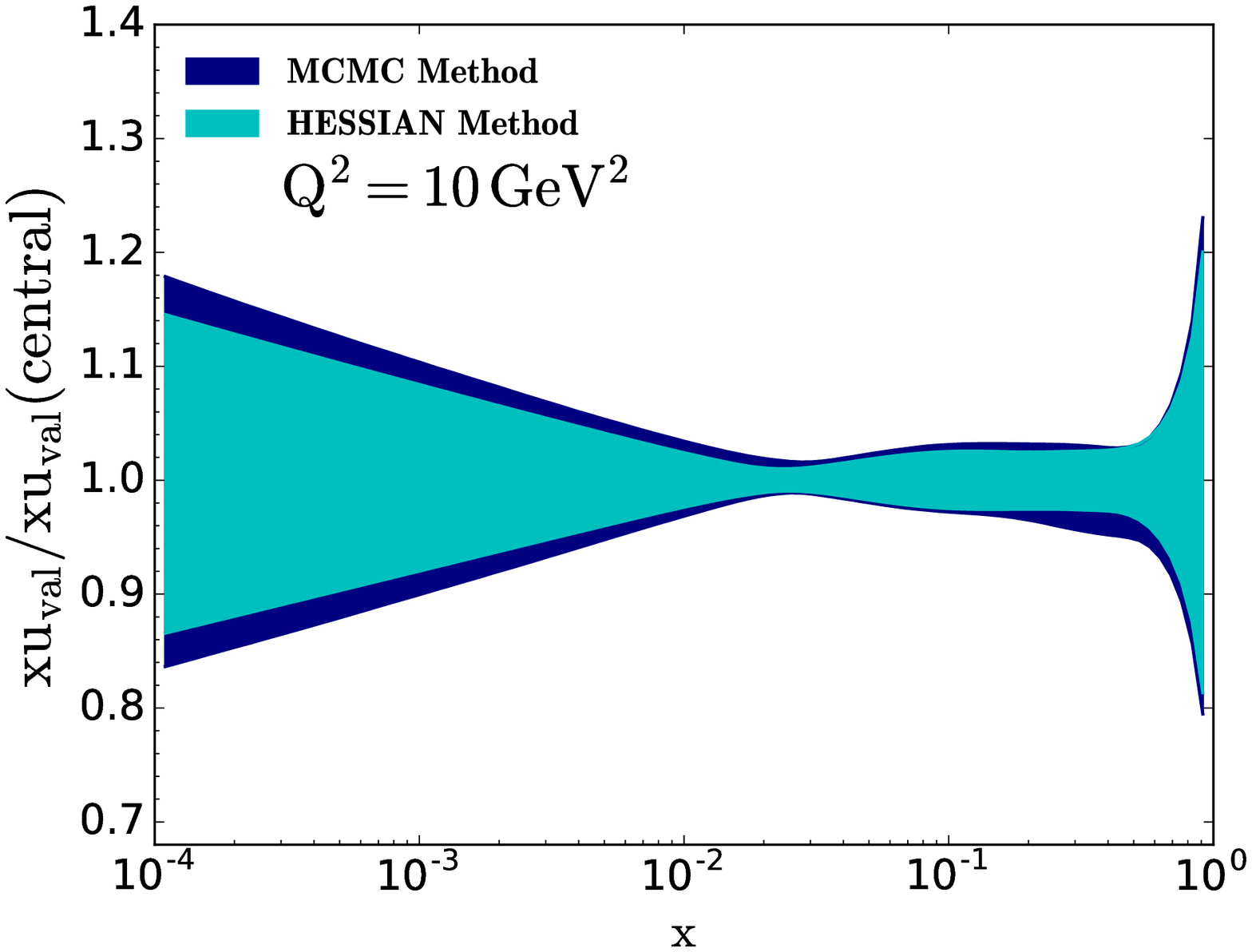}}%\hspace{-0.5cm}
%\raisebox{-0.26mm}
{\includegraphics[width=9cm,height=7.0cm]{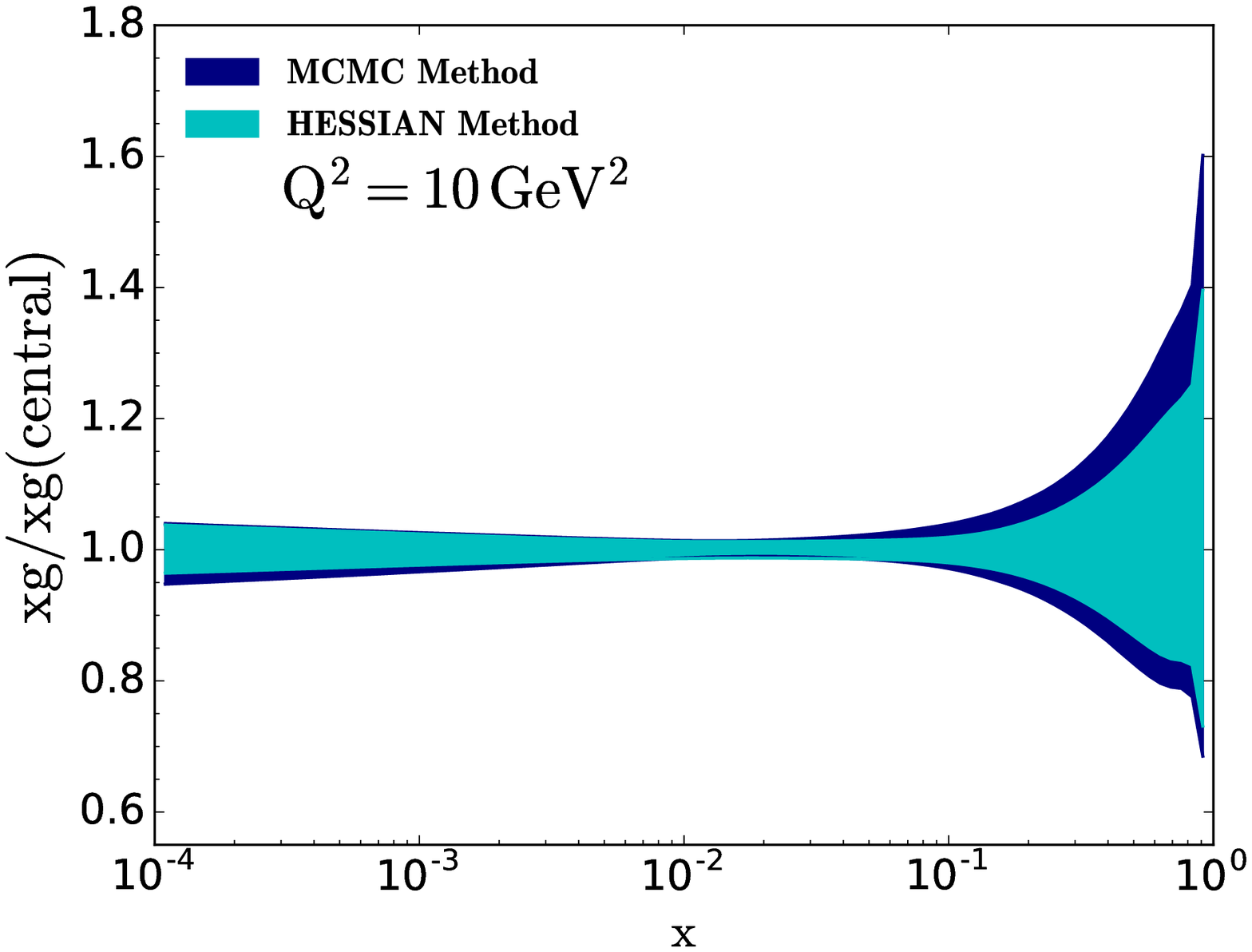}}
\vspace{-0.2cm}
\caption{{\it Comparison of the PDF uncertainties, normalized by the best fit value, as determined by the Hessian and MCMC methods at NLO  for the valence distribution $xu_{val}$ and the gluon distribution $xg$,
at a scale $\mathrm{Q^2}$ = $10\ \mathrm{GeV^2}$.}}
\label{PDfs_uv_Gluon}
\end{figure}

\begin{figure}[H]
\begin{center}
\includegraphics[width=9cm,height=6.0cm]{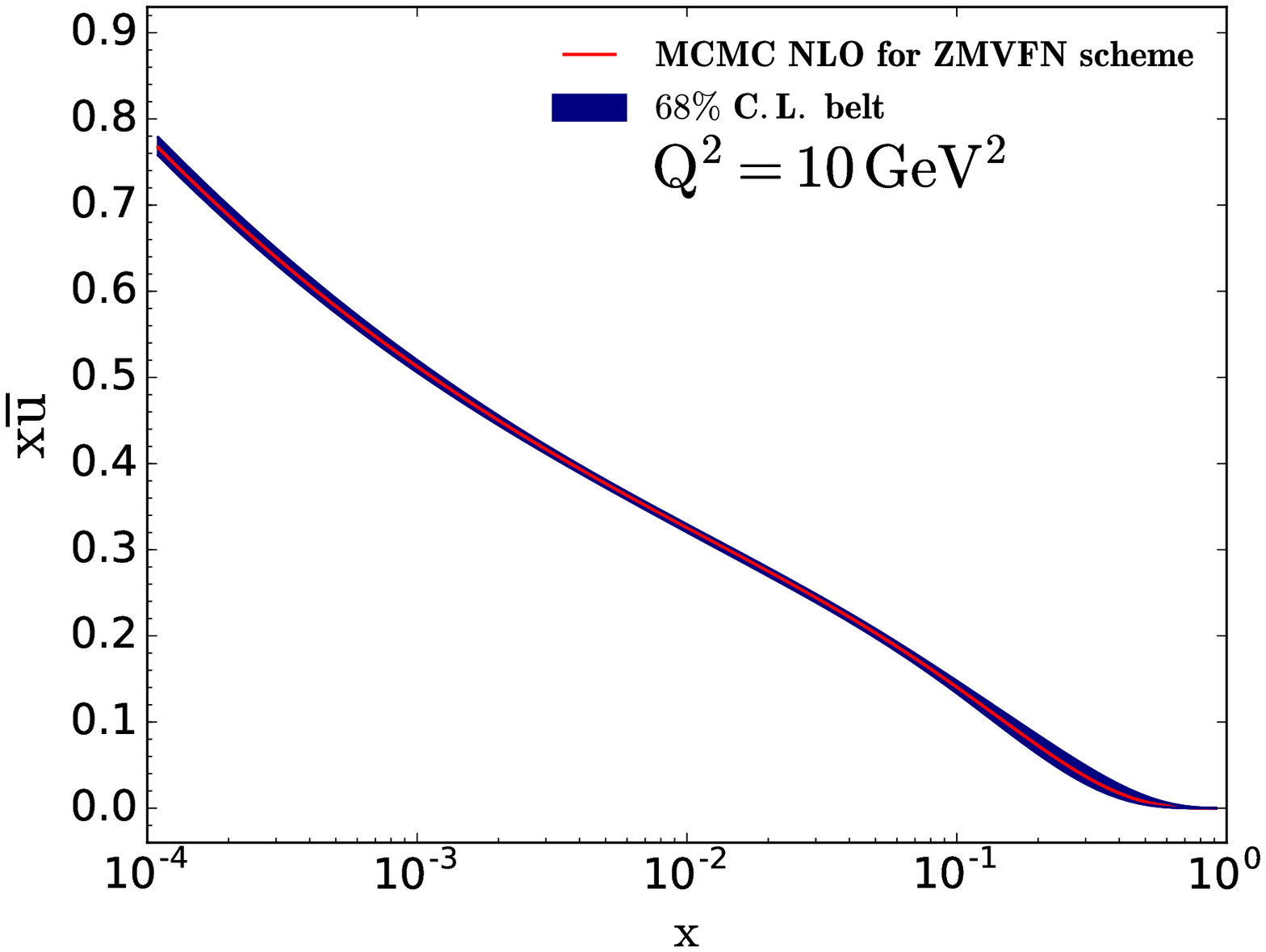}%\hspace{-0.5cm}
\includegraphics[width=9cm,height=6.0cm]{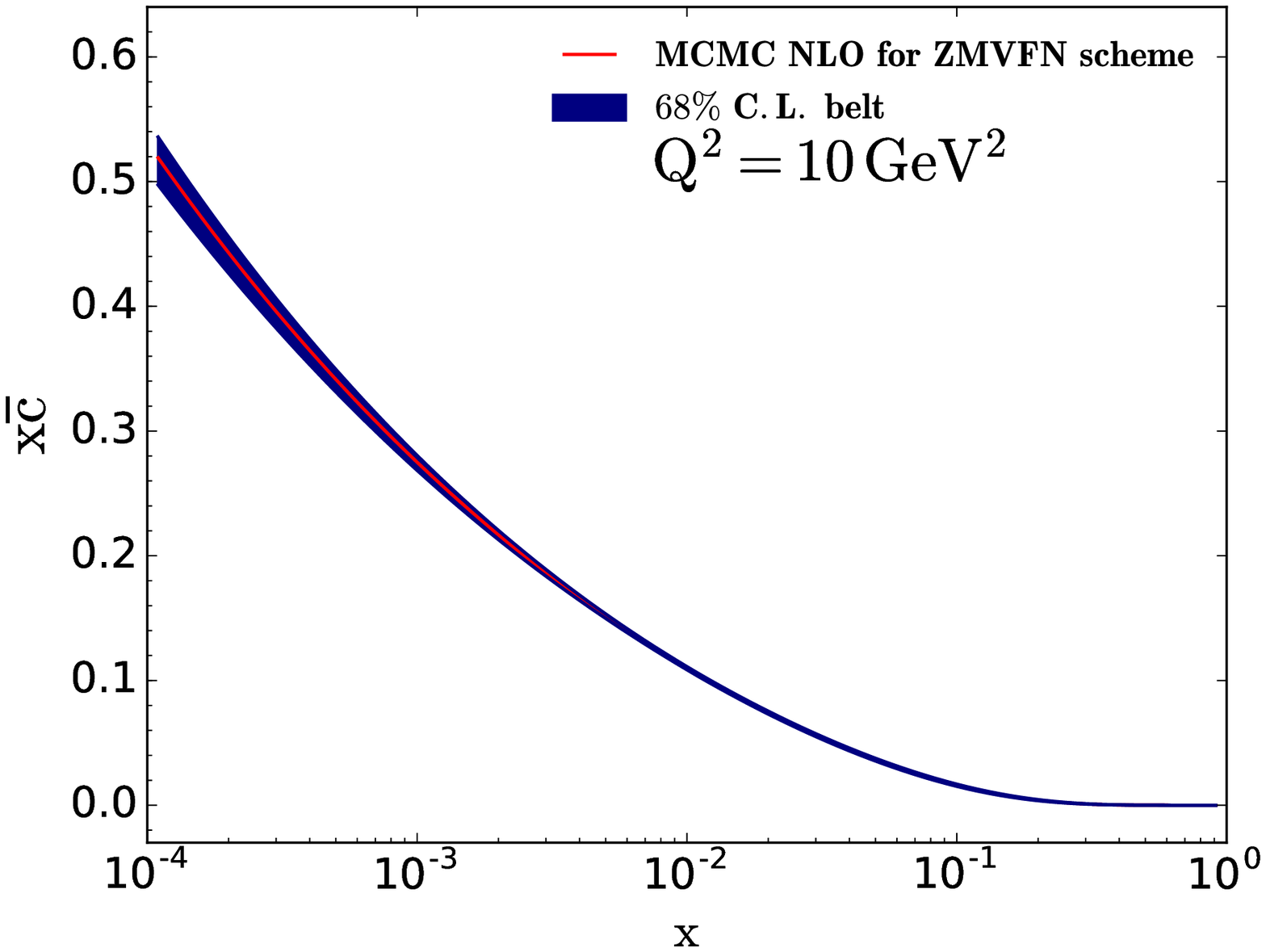}
\includegraphics[width=9cm,height=6.0cm]{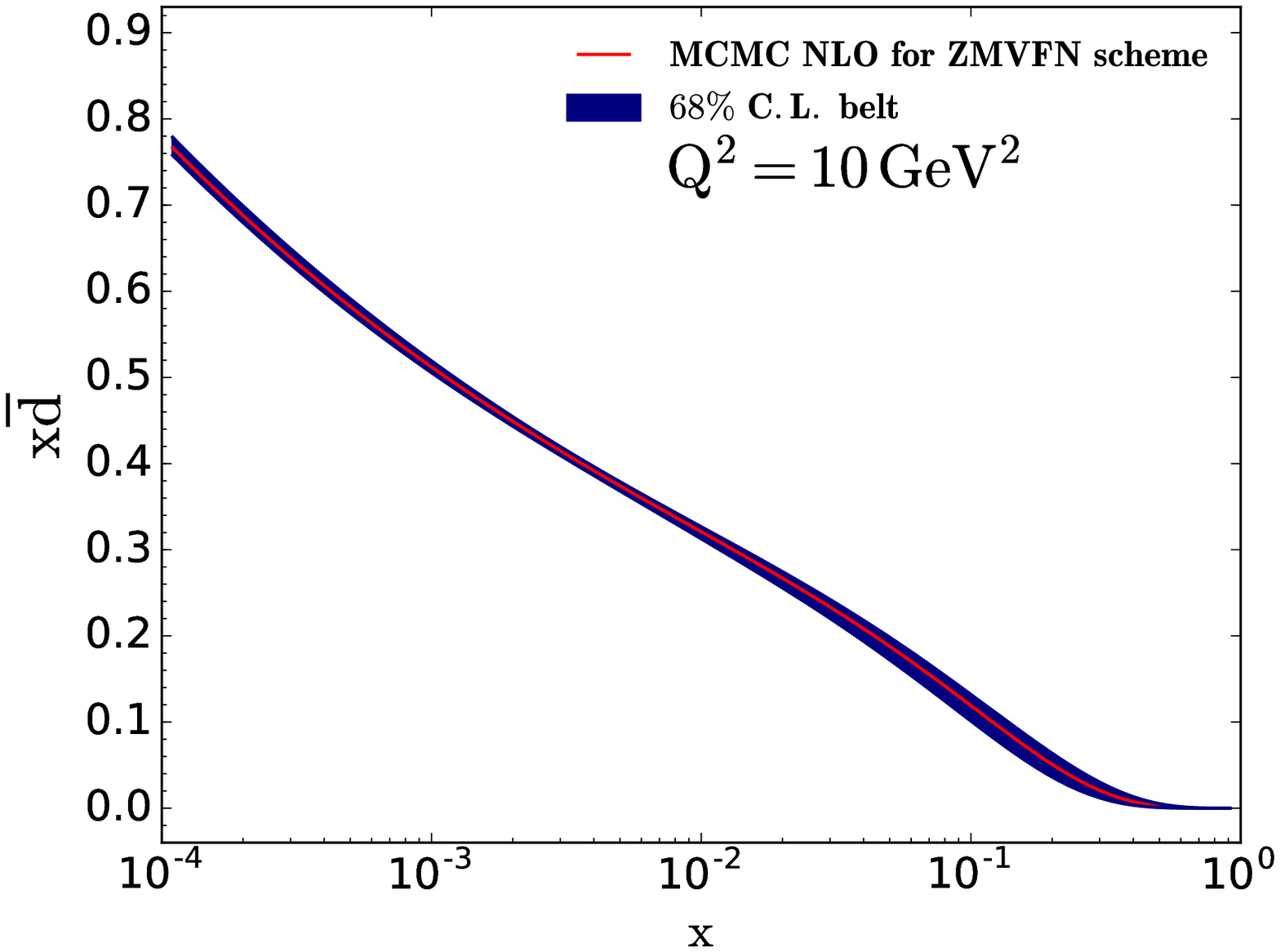}%\hspace{-0.5cm}
\includegraphics[width=9cm,height=6.0cm]{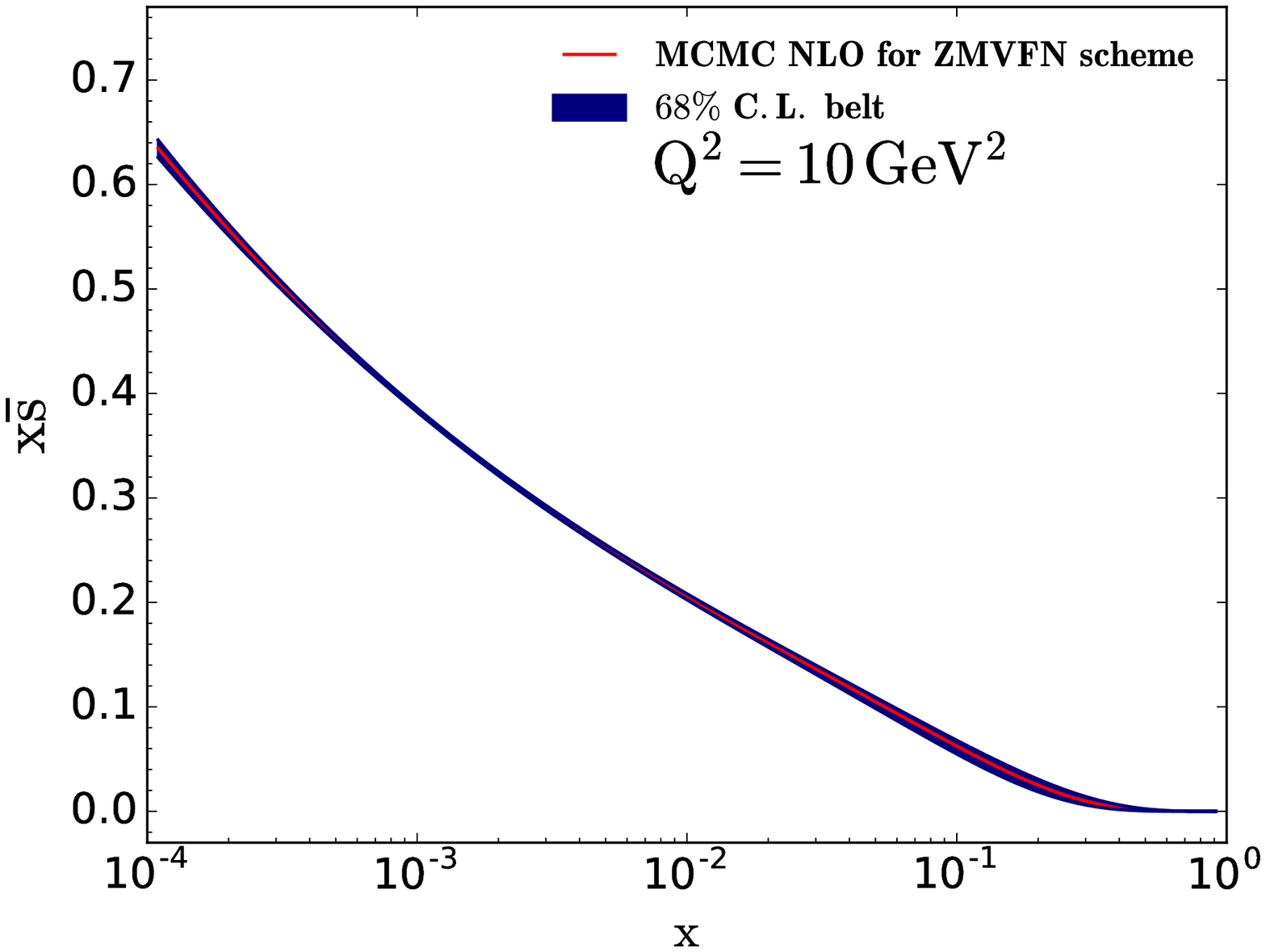}
\end{center}
\vspace{-0.5cm}
\caption{{\it{The MCMC parton distribution functions  $\mathrm{x\overline{u}}$, $\mathrm{x\overline{d}}$, $\mathrm{x\overline{s}}$ and $\mathrm{x\overline{c}}$ at 
$\mathrm{Q^2}$ = $10\ \mathrm{GeV^2}$. The bands show the $68\%$ confidence interval around the best fit value (in solid red line).}}}\label{PDFs_ubar_dbar_sbar_cbar}
\end{figure}

\begin{figure}[H]
\begin{center}
%\vspace{8cm}
%\mbox{\epsfxsize=15cm\epsfysize=7cm\epsffile{hist_HMC_10D_ZMVFNS.eps}}
\includegraphics[width=8.9cm]{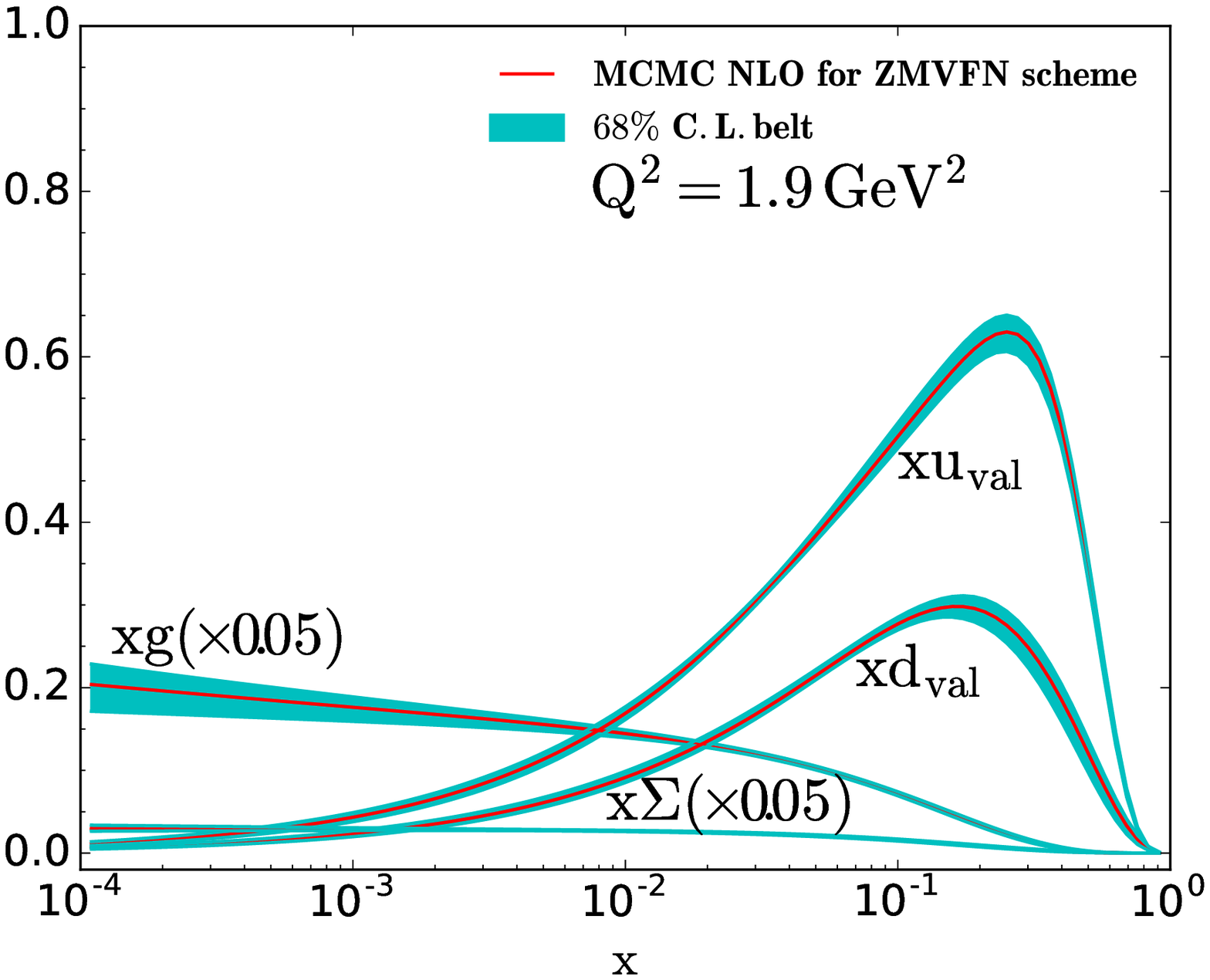}%\hspace{-0.6cm}
\includegraphics[width=8.9cm]{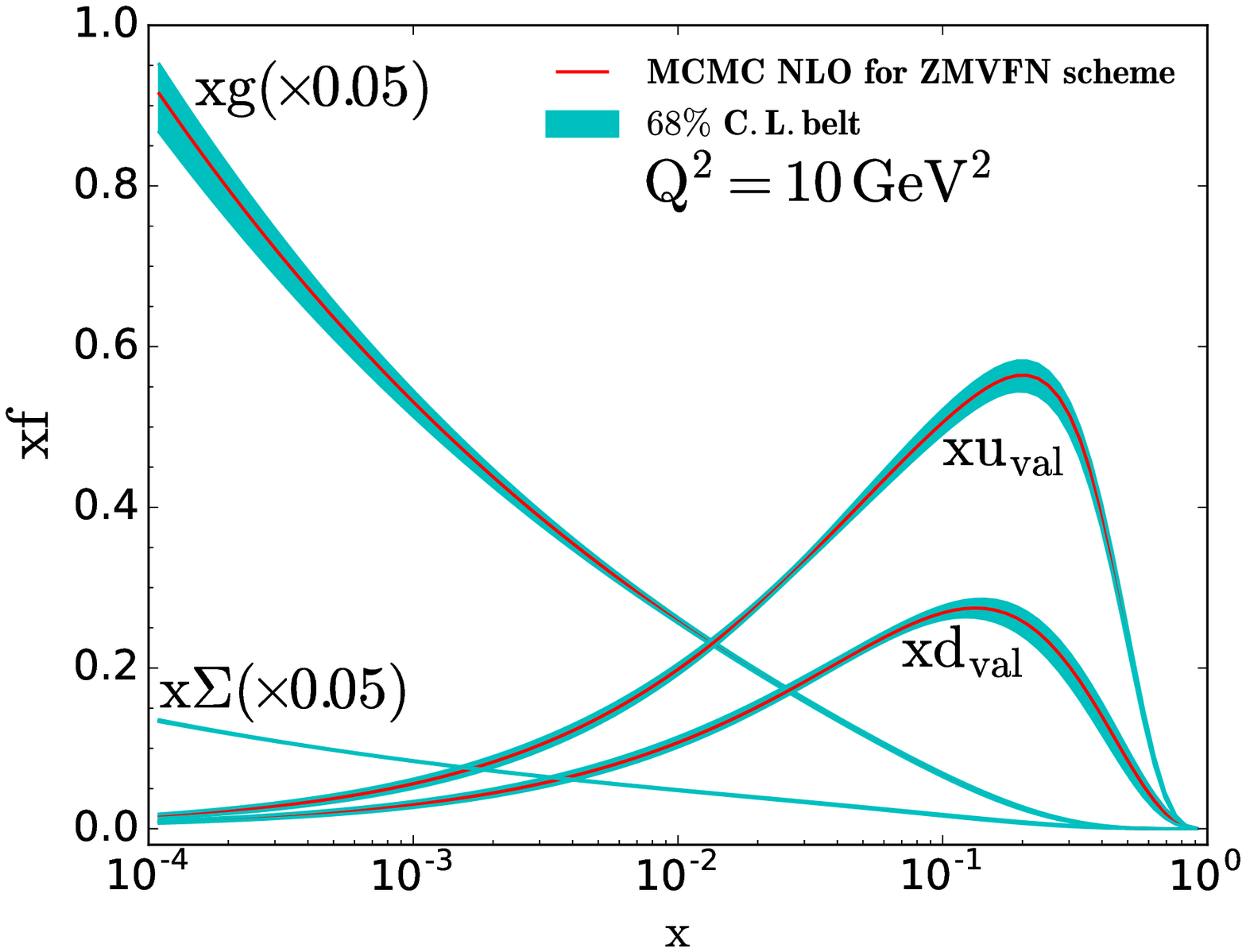}
\end{center}
\vspace{-0.5cm}
\caption{ {\it{PDFs of valence quarks ($xu_{val}$, $xd_{val}$), sea quarks ($x\Sigma = x\overline{u} + x\overline{d} + x\overline{s} + x\overline{c}$, with $x\overline{c}=0$ for $Q^2<m_c^2$), and gluon ($xg$) using MCMC at the scale $\mathrm{Q^2}$ = $1.9\ \mathrm{GeV^2}$ (l.h.s.) and $\mathrm{Q^2}$ = $10\ \mathrm{GeV^2}$ (r.h.s.) obtained with the ZMVFN scheme. The gluon and the sea distributions are scaled down by a factor of 20. 
The experimental uncertainties (68\% confidence limit as defined in the text from the probability density of PDFs) are represented by the green-shaded region. 
%The central value (solid red line)
%is also compared to the HERAPDF1.0 ZMVFNS ones (dashed black line).
}}
}
%\it The parton distribution functions obtained using MCMC4HERAPDF1.0 in the $\mathrm{ZMVFNS\ scheme}$ for $\mathrm{xu}_{\mathrm{val}}$, $\mathrm{xd}_{\mathrm{val}}$, $\mathrm{x\Sigma} = \mathrm{x\overline{u}} + \mathrm{x\overline{d}} + \mathrm{xs} + \mathrm{xc}$\ and\ $\mathrm{xg}$,
%at $\mathrm{Q^2}$ = $10\ \mathrm{GeV^2}$. The bands show
%experimental uncertainties ($68\%$ confidence limits) of the MCMC fit.}}
\label{Fig:Gluon_Sea_uval_dval_pdfs}
\end{figure}

For each $(x,Q^2)$, we determine the $\alpha$\%-confidence interval around the best fit value of the PDF (with typically $\alpha=68$ or $\alpha=95$) by considering the region of the distribution on each side of the best fit, and taking 
$\alpha$\% of the data on each of these regions.  
This provides the (non necessarily symmetric)  $\alpha$\%-confidence limit envelops we show in Figures \ref{PDfs_uval_Gluon} to \ref{Fig:Gluon_Sea_uval_dval_pdfs}. 

The gluon and $u_{val}$ parton distributions obtained this way are plotted as functions of $x$ and for $Q^2=10$ GeV$^2$ in Figure \ref{PDfs_uval_Gluon}. The central PDF is the best fit value. These MCMC PDFs are compared with the HERAPDF1.0 PDFs (ZMVFN scheme), with a direct
comparison in Figure \ref{PDfs_uval_Gluon}, and a ratio plot in Figure \ref{Fig:Ratio_Gluon_Sea_uval_dval_pdfs}. 
They are, as expected, very close, both in central value and in confidence interval. Maximum likelihood estimator and least square method are indeed equivalent under Gaussian assumption, which in our case can be reasonably applied, as mentioned in the previous section. 

%These remarks also hold for all other PDFs, though we plotted in Figure \ref{PDfs_uval_Gluon} only comparison between
%MCMC and minimization for gluon and $u_{val}$ ones. 

The uncertainties obtained by the MCMC method and the Hessian method are also consistent within the kinematic range of HERA. This is demonstrated in figure \ref{PDfs_uv_Gluon} where experimental
uncertainties -- normalized by the best fit value --  obtained for HERAPDF1.0 NLO and MCMC NLO respectively by the Hessian and MCMC methods are
compared for the $u_{val}$ and the gluon distributions. The MCMC uncertainties tend to be slightly larger than the standard deviations obtained in
the Hessian approach.
For completeness, we also display in Figure \ref{PDFs_ubar_dbar_sbar_cbar} the anti-quark 
PDFs at $Q^2=10$ GeV$^2$, and in Figure \ref{Fig:Gluon_Sea_uval_dval_pdfs} the central value and 68\% confidence
limit interval for $xu_{val}$, $xd_{val}$, $xg$ and 
$x\Sigma$ ($x\Sigma = x\overline{u} + x\overline{d} + x\overline{s} + x\overline{c}$, with $x\overline{c}=0$ for $Q^2<m_c^2$) on the same plot, at $Q^2=1.9$ GeV$^2$ and $Q^2=10$ GeV$^2$. 

%\textcolor{green}{In figure \ref{Fig:Gluon_Sea_uval_dval_pdfs}, as we go from $\mathrm{Q^2}=1.9~\mathrm{GeV^2}$ (left) to $\mathrm{Q^2}=10~\mathrm{GeV^2}$(right), the valence quarks distributions $xu_{val}$ and $xd_{val}$ decrease at large $x$ while the gluon distribution $xg$ and sea distribution $x\Sigma$ increase at small $x$ : it is the effect of the DGLAP evolution.}

\section{Conclusion and outlook}

We have shown that MCMC, known to be well suited to multi-parameter determination, is applicable to PDFs determination and that Bayesian parameter inference approach applied to global PDFs analysis can lead to a deeper insight into PDFs uncertainties. The innovative procedure we implemented, which combines Monte Carlo technics, lattice-developed algorithms and global PDFs analysis is complementary to the existing methods. 
We have for the first time applied Hybrid Monte Carlo algorithm to PDFs determination and computed marginal probability densities of PDFs parameters, and PDFs themselves. This
allows to study the probability distribution of these functions, to determine mean, most probable 
and median values, and to extract confidence intervals in a statistically controlled way.

This work will lead to an innovative PDFs uncertainties determination, and thus to reliable determination of uncertainties for many collider observables, in a way complementary to the existing
methods. This will also open new ways to analyze the impact of new data sets being added to the analysis, to check which dataset has outliers and if these latter can be tolerated. 

This feasibility study paves the way for a more complete PDFs determination by MCMC technics, and our goal is to extend the present work to the full ensemble of PDFs free parameters, including also as parameters, the strong coupling constant and $c$ and $b$ quark masses. We will consider
more complex $\chi^2$ functions including correlation and complete our analysis on a fully realistic case, studying in particular the impact of priors. No doubt that Markov Chain Monte Carlo 
methods will give interesting and valuable informations on PDFs and will contributed to our deeper understanding of these key elements of QCD.

%%%%%%%% Acknowledgments %%%%%%%%%%%%%%%%%

\section{Acknowledgments}

We thank L. Derome and D. Maurin for fruitful discussions, and the xFITTER support team for
their help. We are also grateful to the CFM Foundation for Research for its financial support. 
Most of the computation has been done in Lyon-CCIN2P3, thanks to the CPUs resources provided. We are grateful to the staff members of this Computing Center for their constant support.

%%%%%%%%%%%%%%%%%%%%%%%%%%%%%%%%%%%%%%%%%%%%%%%%%%%%%%%%%%%%%%%%%%%%

\begin{thebibliography}{100}
\bibitem{PDF_LQCD}X. Ji, %{\it{Parton Physics on a Euclidean Lattice}},
Phys.Rev.Lett. {\bf{110}}, 262002 (2013). 
%arXiv:1305.1539 [hep-ph].
\bibitem{Lai_PRD97}H. L. Lai, J. Huston, S. Kuhlmann, F. Olness, J. Owens, D. Soper, W. K. Tung, and H. Weerts, Phys. Rev. {\bf{D55}}, 1280 (1997). 
\bibitem{Martin_DTP96}A. D. Martin, R. G. Roberts, M. G. Ryskin, W. J. Stirling,  
%DTP-96102, hep-ph/9612449 (1996); 
Eur. Phys. J. {\bf{C2}}, 287 (1998);
A. D. Martin, R. G. Roberts, W. J. Stirling and R. S. Thorne, Eur. Phys. J. {\bf{C 28}}, 455
(2003); %[hep-ph/0211080]; 
A.D. Martin, R.G. Roberts, W.J. Stirling, R.S. Thorne,
Eur. Phys. J. {\bf{C35}}, 325 (2004).
% arXiv:hep-ph/0308087 (2003).
\bibitem{Giele_98}Walter T. Giele and Stephane Keller, Phys. Rev. {\bf{D58}}, 094023 (1998).
%hep-ph/9803393, FERMILAB-Pub98/082-T (1998).
\bibitem{Huston_96}J. Huston, E. Kovacs, S. Kuhlmann, H. L. Lai, J. F. Owens, D. Soper, and W. K. Tung, Phys. Rev. Lett. {\bf{77}}, 444 (1996).
\bibitem{J_Pumplin_PRD2001}	
%{\it{Uncertainties of predictions from parton distribution functions. 2. The Hessian method}},
J. Pumplin, D. Stump, R. Brock, D. Casey, J. Huston, J. Kalk, H.L. Lai, W.K. Tung, Phys. Rev.  {\bf{D65}}, 014013 (2001); %e-Print: hep-ph/0101032
J. Pumplin, D. R. Stump and W. K. Tung, Phys. Rev. {\bf{D65}}, 014011 (2001); %[hep-ph/0008191];
D. Stump, J. Pumplin, R. Brock, D. Casey, J. Huston, J. Kalk, H.L. Lai, W. K. Tung,
Phys. Rev. {\bf{D65}}, 014012 (2001); %[hep-ph/0101051]
\bibitem{D_Stump_PRD2001}
%{\it{Uncertainties of predictions from parton distribution functions. 1. The Lagrange multiplier method}},
D. Stump, J. Pumplin, R. Brock, D. Casey, J. Huston, J. Kalk, H.L. Lai, W.K. Tung, Phys. Rev.  {\bf{D65}}, 014012 (2001). % e-Print: hep-ph/0101051 
\bibitem{Giele_2001}W. T. Giele and S. Keller, Phys. Rev. {\bf{D58}},
094023 (1998): %[hep-ph/9803393]
W. T. Giele, S. Keller and D. Kosower, FERMILAB-PUB-01-498-T, arXiv:0104052 [hep-ph]  
(2001).
%{\it{Parton distribution function uncertainties}}hep-ph/0104052
\bibitem{Gilks}W.R. Gilks, S. Richardson and D.J. Speigelhalter, 
{\it{Markov Chain Monte Carlo in practice}}, Chapman \& Hall/CRC Interdisciplinary Statistics ed., ISBN 9780412055515.
\bibitem{Sokal_1989}A. D. Sokal, {\it{Monte Carlo Methods in statistical machanics: foundations and new algorithms}}, Cours de Troisi\`eme Cycle de la Physique en Suisse Romande (Lausanne, Switzerland) (1989). 
\bibitem{DelDebbio:2007ee} 
L. Del Debbio, S. Forte, J. I. Latorre, A. Piccione, and J. Rojo [NNPDF Collaboration],
%{\it{Neural network determination of parton distributions: The Nonsinglet case}},
JHEP {\bf 0703}, 039 (2007); %,  [hep-ph/0701127]; 
J.~C.~Rojo,
%{\it{The Neural network approach to parton distribution functions,}}, 
PhD thesis, hep-ph/0607122 (2006).
\bibitem{Nikolaev_1991}N. N. Nikolaev and B. Zakharov, Z. Phys. {\bf{C49}}, 607 (1991); A. H. Mueller, Nucl. Phys. {\bf{B 415}}, 373 (1994).
\bibitem{Collins_2011} J. Collins,  {\it{Foundations of perturbative QCD}}, vol. 32
(Cambridge monographs on particle physics, nuclear physics and cosmology, 2011). 
\bibitem{Aybat_2011}S. M. Aybat and T. C. Rogers, Phys. Rev. {\bf{D83}}, 114042
(2011); %[arXiv:1101.5057]; 
M. G. A. Buffing, A. Mukherjee, and P. J. Mulders, Phys. Rev. 
D86, 074030 (2012): %[arXiv:1207.3221]; 
P. Mulders, Pramana {\bf{72}}, 83 (2009) arXiv:0806.1134 [hep-ph]; 
S. Jadach and M. Skrzypek, Acta Phys. Polon. {\bf{B40}},
2071 (2009); % [arXiv:0905.1399]; 
F. Hautmann, Acta Phys. Polon. {\bf{B40}}, 2139 (2009).
\bibitem{MCMC_Neal}R. M. Neal in {\it{Handbook of Markov Chain Monte Carlo}}, S. Brooks, A. Gelman, G. L. Jones and X. Meng Ed., Chapman and Hall/CRC Press (2011).
\bibitem{Metropolis}
N. Metropolis, A. W. Rosenbluth, M. N. Rosenbluth, A. H. Teller, and E. Teller,  
%{\it{Equations of state calculations by fast computing machine.}}, 
J. Chem Phys. {\bf{21}}, 1087 (1953).
\bibitem{Hastings}W. K. Hastings, %{\it{Monte Carlo sampling methods using Markov chains and their applications.}}
Biometrica, {\bf{57}}, 97 (1970).
\bibitem{deBerg_2000} M. de Berg, O. Cheong. M. van Kreveld and  M. Overmars, 
{\it{Computational Geometry - Algorithms and Applications}} (Springer-Verlag),  2nd revised edition, ISBN: 978-3-662-04247-2.
\bibitem{HMC_Duane1987}S. Duane, A. D. Kennedy, B. J. Pendleton and D. Roweth, Phys. Lett.
{\bf{B195}}, 216 (1987).
\bibitem{DeromePutze_2014}A. Putze and L. Derome, 
%{\it{The Grenoble Analysis Toolkit (GreAT)"A statistical analysis framework}} 
Physics of the Dark Universe {\bf{Vol. 5}}, 29 (2014).
\bibitem{Putze}A. Putze, L. Derome, D. Maurin, L. Perotto, and R. Taillet,
%{ \it{A Markov Chain Monte Carlo technique to sample transport and source parameters of Galactic cosmic rays}} 
Astron. Astrophys.  {\bf{Vol. 497}} No 3, 991 (2009).
\bibitem{Wolff_2003}U. Wolff, 
%{ \it{Monte Carlo errors with less errors}}
Comput. Phys. Commun. {\bf{156}},
143 (2004).% , hep-lat/0306017.
\bibitem{Quenouille_1956}M. H. Quenouille, 
%{ \it{Notes on biais in estimation}}, 
Biometrika, {\bf{Vol. 43}}, No 3/4, 353 (1956).
\bibitem{HeraFitter}S. Alekhin et al., Eur. Phys. J. {\bf{C75}} No 7, 304 (2015), http://www.xfitter.org/;
%{\it{HERAFitter, Open Source QCD Fit Project}}, 
%arXiv:1410.4412v3 [hep-ph], Desy Report 14-188,  (2015).
%2) "Combined Measurement and QCD Analysis of the Inclusive e+- p Scattering Cross Sections at HERA."
H1 and ZEUS Collaboration (F.D. Aaron et al.), J.H.E.P. {\bf{1001}}, 109 (2010).
%e-Print: arXiv:0911.0884 [hep-ex]
%3) "A Precision Measurement of the Inclusive ep Scattering Cross Section at HERA."
H1 Collaboration (F.D. Aaron et al.),  Eur. Phys. J. {\bf{C64}}, 561 (2009); 
%e-Print: arXiv:0904.3513 [hep-ex]
%QCDNUM  ( evolution code )
%-------
%"Fast QCD Evolution and Convolution", 
M. Botje, Comput. Phys. Commun. {\bf{182}}, 490 (2011); 
%e-Print: arXiv:1005.1481 [hep-ph]
%MINUIT  ( minimisation code )
%-------
F. James, M. Roos (CERN), Comput. Phys. Commun. {\bf{10}}, 343 (1975);
%"Measurement and QCD analysis of neutral and charged current cross-sections at HERA"
H1 Collaboration (C. Adloff et al.), Eur. Phys. J. {\bf{C30}},1 (2003);
%e-Print: hep-ex/0304003
%"A ZEUS next-to-leading-order QCD analysis of data on deep inelastic scattering"
ZEUS Collaboration (S. Chekanov et al.), Phys. Rev. {\bf{D67}}, 012007(2003);
%e-Print: hep-ex/0208023
%
%'HERAPDF':
%Add two references above (for 'H12000'), in addition:
%"Measurement of the Inclusive ep Scattering Cross Section at Low Q^2 and x at HERA"
F.D. Aaron et al., Eur. Phys. J. {\bf{C63}}, 625 (2009);
%e-Print: arXiv:0904.0929 [hep-ex]
%"An NLO QCD analysis of inclusive cross-section and jet-production data from the zeus experiment"
ZEUS Collaboration (S. Chekanov et al.), Eur. Phys. J. {\bf{C42}}, 1 (2005);
%e-Print: hep-ph/0503274
%"New generation of parton distributions with uncertainties from global QCD analysis"
J. Pumplin, D.R. Stump, J. Huston, H.L. Lai, P. M. Nadolsky, W.K. Tung, J.H.E.P. {\bf{0207}}, 012 (2002).
%e-Print: hep-ph/0201195
\end{thebibliography}
\end{document}